\documentclass[fp,twocolumn]{jpsj3}
\usepackage{graphicx,color}

\def\gsim{\,$\raise0.3ex\hbox{$>$}\llap{\lower0.8ex\hbox{$\sim$}}$\,}
\def\lsim{\,$\raise0.3ex\hbox{$<$}\llap{\lower0.8ex\hbox{$\sim$}}$\,}
\def\cH{{\cal H}}
\def\rr{{\rm r}}
\def\rl{{\rm l}}
\def\up{\uparrow}
\def\dn{\downarrow}
\def\cHeff{{\mathcal H}_{\rm eff}}
\def\Jeff{J_{\rm eff}}
\def\Deltaeff{\Delta_{\rm eff}}
\def\Deff{D_{\rm eff}}
\def\ket#1{|#1\rangle} 

\title{Ground-State Phase Diagram of an Anisotropic S=1/2 Ladder\\
with Different Leg Interactions}

\author{Takashi Tonegawa$^{1,2}$\thanks{E-mail address:
tone0115@vivid.ocn.ne.jp}, Toshiya Hikihara$^{3}$,
Kiyomi Okamoto$^4$, Shunsuke C. Furuya$^5$, and
T\^oru Sakai$^{6,7}$
}

\inst{$^1$Professor Emeritus, Kobe University, Kobe 657-8501, Japan\\
$^{2}$Department of Physical Science, Osaka Prefecture University,
Sakai 599-8531, Japan\\
$^{3}$Faculty of Science and Technology, Gunma University, Kiryu
376-8515, Japan\\
$^{4}$College of Engineering, Shibaura Institute of Technology,
Saitama 337-8570, Japan\\
$^{5}$Condensed Matter Theory Laboratory, RIKEN, Wako 351-0198,
Japan\\
$^{6}$Graduate School of Material Science, University of Hyogo, Hyogo
678-1297, Japan\\
$^{7}$National Institutes for Quantum and Radiological Science and
Technology (QST), SPring-8, Hyogo 679-5148, Japan
}

\abst{
We explore the ground-state phase diagram of the \hbox{$S\!=\!1/2$} two-leg
ladder with different leg interactions.  The $xy$- and $z$-components of the
leg interactions between nearest-neighbor spins in the $a$ ($b$) leg are
respectively denoted by $J_{{\rm l},a}$ and
$\Delta_{\rm l} J_{{\rm l},a}$ ($J_{{\rm l},b}$ and
$\Delta_{\rm l} J_{{\rm l},b}$).  On the
other hand, the $xy$- and $z$-components of the uniform rung interactions are
respectively denoted by $\Gamma_{\rm r} J_{{\rm r}}$ and $J_{{\rm r}}$.  In
the above, $\Delta_{\rm l}$ and $\Gamma_{\rm r}$ are the $XXZ$-type anisotropy
parameters for the leg and rung interactions, respectively.  This system
has frustration when \hbox{$J_{{\rm l},a} J_{{\rm l},b}\!<\!0$} irrespective
of the sign of $J_{\rm r}$.  The phase diagram on
the $\Delta_{\rm l}$ (\hbox{$|\Delta_{\rm l}|\!\leq\!1.0$}) versus
$J_{{\rm l},b}$ (\hbox{$-2.0\!\leq\!J_{{\rm l},b}\!\leq\!3.0$}) plane in the
case where \hbox{$J_{{\rm l},a}\!=\!0.2$}, \hbox{$J_{{\rm r}}\!=\!-1.0$},
and \hbox{$\Gamma_{\rm r}\!=\!0.5$} is determined numerically.  We employ
physical considerations and perform level spectroscopy and phenomenological
renormalization-group analyses of the numerical data obtained by the exact
diagonalization method.  The resultant phase diagram contains the
ferromagnetic, Haldane, N{\'e}el, nematic Tomonaga--Luttinger liquid (TLL),
partial ferrimagnetic, and $XY1$ phases.  Interestingly enough, the nematic
TLL phase appears in the strong-rung unfrustrated region as well as in the
strong-rung frustrated region.  We perform first-order perturbational
calculations from the strong-rung coupling limit to elucidate the
characteristic features of the phase diagram.  Furthermore, we carry out
density-matrix renormalization-group calculations for some physical quantities
such as the energy gaps, the local magnetization, and the spin correlation
functions to supplement the reliability of the phase diagram.  The phase
diagram on the $\Gamma_{\rm r}$ (\hbox{$0.0\!\leq\!\Gamma_{\rm r}\!\leq\!1.0$})
versus $J_{{\rm l},b}$ (\hbox{$-1.0\!\leq\!J_{{\rm l},b}\!\leq\!2.0$}) plane
in the case where \hbox{$J_{{\rm l},a}\!=\!0.2$},
\hbox{$J_{{\rm r}}\!=\!-1.0$}, and \hbox{$\Delta_{\rm l}\!=\!1.0$} is also
discussed briefly.
}

\kword{ground-state phase diagram, anisotropic \hbox{$S\!=\!1/2$} ladder,
different leg interactions, nematic Tomonaga-Luttinger liquid phase,
exact diagonalization calculation, perturbational calculation, density-matrix
renormalization-group calculation}

\begin{document}
\maketitle

\section{Introduction}

Over the past several decades, a great deal of effort has been devoted to
clarifying the frustration effect on the ground-state properties of
low-dimensional quantum spin systems with competing interactions.  It is now
well known experimentally as well as theoretically and numerically that a
reciprocal influence between the strong quantum fluctuation and the
geometrical frustration in these systems induces various exotic quantum ground
states such as the dimer state accompanying spontaneous translational symmetry
breaking, the spin-nematic Tomonaga-Luttinger liquid (TLL) state, and so on.
The former state is realized in an \hbox{$S\!=\!1/2$} zigzag chain with
antiferromagnetic nearest-neighbor (nn) and next-nearest-neighbor (nnn)
interactions.~\cite{majumdar-ghosh,tonegawa-harada,okamoto-nomura,
tonegawa-harada-kaburagi,nomura-okamoto-1,nomura-okamoto-2}  On the
other hand, the latter state is realized in an anisotropic \hbox{$S\!=\!1$}
chain,~\cite{schulz,nijs-rommelse,chen-etal} in an \hbox{$S\!=\!1$} chain with
bilinear and biquadratic interactions,~\cite{lauchli-etal} and in an
\hbox{$S\!=\!1/2$} zigzag chain with ferromagnetic nn and antiferromagnetic nnn
interactions under an external magnetic
field.~\cite{vekua-etal,hikihara-etal-1,sudan-etal}

In the case of an \hbox{$S\!=\!1/2$} two-leg ladder system, the frustration is
usually introduced by adding nnn leg and/or diagonal interactions to an
original system with nn leg and rung interactions, as has been extensively
investigated.~\cite{frustrated-ladder-1,frustrated-ladder-2,
frustrated-ladder-3}  As another example, an \hbox{$S\!=\!1/2$} ladder with
uniform nn leg and alternating rung interactions has also been studied by
several authors.~\cite{ladder-altrung-1,ladder-altrung-2,ladder-altrung-tone}
We~\cite{ladder-altrung-tone} have discussed the ground-state phase diagram
in the frustrated case where rung interactions are
ferromagnetically-antiferromagnetically alternating and have a common
Ising-type anisotropy, while leg interactions are antiferromagnetically uniform
and isotropic.  Our results show that, when the leg interactions are relatively
weak compared with the rung interactions, the incommensurate Haldane state as
well as the commensurate one appears as the ground state in the whole range of
the Ising-type anisotropy parameter.  This appearance of the Haldane state in
the case where the Ising character of rung interactions is strong is contrary
to the ordinary situation, and is called the inversion phenomenon concerning
the interaction anisotropy~\cite{inv-1,inv-2,inv-3,inv-4}.
\begin{figure}[t]
   \begin{center}
       \scalebox{0.6}{\includegraphics{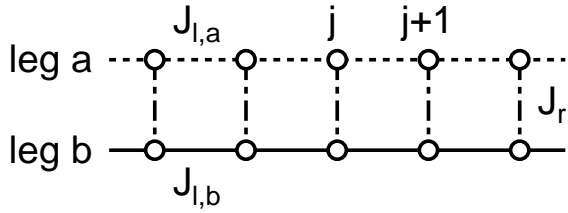}}
   \end{center}
       \caption{Schematic sketch of the \hbox{$S\!=\!1/2$} ladder with
                different leg interactions.  Open circles denote
                \hbox{$S\!=\!1/2$} spins, while solid, dotted and dot-dashed
                lines denote, respectively, the interaction constants,
                $J_{{\rm l},a}$, $J_{{\rm l},b}$, and $J_{\rm r}$.
                }
       \label{fig:model}
\end{figure}

In the present paper, we explore the ground-state phase diagram of an
anisotropic \hbox{$S\!=\!1/2$} ladder with different leg interactions, which
is schematically sketched in Fig.~\ref{fig:model} and described by the
Hamiltonian
\begin{eqnarray}
{\cal H}
    \!\!\!\!\!\!&=&\!\!\!\!\!\!
           J_{{\rm l},a} \sum_{j=1}^{L}
                   \left\{S_{j,a}^x S_{j+1,a}^x + S_{j,a}^y S_{j+1,a}^y
                   + \Delta_{\rm l} S_{j,a}^z S_{j+1,a}^z \right\}
                                                                  \nonumber\\
      &&\!\!\!\!\!\!\!\! + \,J_{{\rm l},b} \sum_{j=1}^{L}
                   \left\{S_{j,b}^x S_{j+1,b}^x + S_{j,b}^y S_{j+1,b}^y
                   + \Delta_{\rm l} S_{j,b}^z S_{j+1,b}^z \right\}
                                                                  \nonumber\\
      &&\!\!\!\!\!\!\!\! + \,J_{\rm r} \sum_{j=1}^{L}
             \left\{\Gamma_{\rm r}
                      \left(S_{j,a}^x S_{j,b}^x + S_{j,a}^y S_{j,b}^y\right)
                                        + S_{j,a}^z S_{j,b}^z \right\}\,.
\label{eq:hamiltonian}
\end{eqnarray}
Here, $S_{j,\ell}^x$, $S_{j,\ell}^y$, and $S_{j,\ell}^z$ are, respectively, the
$x$-, $y$-, and $z$-components of the \hbox{$S\!=\!1/2$} operator
${\vec S}_{j,\ell}$ at the \hbox{$(j,\ell)$} site assigned by the $j$th rung
and the $\ell(=\!a~{\rm or}~b)$
leg; $J_{{\rm l},a}$ and $J_{{\rm l},b}$ denote, respectively, the magnitudes
of the $a$ leg and $b$ leg interactions, while $J_{\rm r}$ denotes that of the
rung interaction; $\Delta_{\rm l}$ and $\Gamma_{\rm r}$ are, respectively, the
parameters representing the  $XXZ$-type anisotropies of the former and latter
interactions; $L$ is the total number of rungs, which is assumed to be even.
It is emphasized that this system has frustration when
\hbox{$J_{{\rm l},a} J_{{\rm l},b}\!<\!0$} irrespective of the sign of
$J_{\rm r}$.

It is unfortunate that real materials for which the above ${\cal H}$ is a good
model Hamiltonian have been neither synthesized nor found so far.  We do not
think, however, that it is physically unrealistic.  Actually, Yamaguchi
{\it et} {\it al.}~\cite{yamaguchi-etal-1,yamaguchi-etal-2} have recently
demonstrated the modulation of magnetic interactions in spin ladder systems by
using verdazyl-radical crystals.  The flexibility of molecular arrangements in
such organic-radical materials is expected to realize \hbox{$S\!=\!1/2$}
ladder systems with different leg interactions.

Our Hamiltonian ${\cal H}$ contains five interaction parameters.  Throughout
the following discussions, we focus our attention upon the case where
\hbox{$J_{{\rm l},a}\!=\!0.2$}, \hbox{$-2.0\!\leq\!J_{{\rm l},b}\!\leq\!3.0$},
\hbox{$J_{{\rm r}}\!=\!-1.0$}, \hbox{$|\Delta_{\rm l}|\!\leq\!1.0$},
and \hbox{$\Gamma_{\rm r}\!=\!0.5$}, unless otherwise stated.  Here, we
choose $\vert J_{{\rm r}}\vert$ as the unit of energy.  It is noted that the
anisotropies of the leg and rung interactions are, respectively, of the $XY$
and Ising types.  The motivation for treating this case is as follows.  When
the ferromagnetic rung interactions with Ising-type anisotropy are much
stronger than both kinds of leg
interactions (\hbox{$\vert J_{{\rm l},b}\vert\!\ll\!1.0$}), a pair of
\hbox{$S\!=\!1/2$} spins at each rung forms a bound state of two magnons with
\hbox{$S_{j,a}^z\!+\!S_{j,b}^z\!=\!\pm 1$}.  This may lead to the nematic
TLL state, which accompanies two-magnon bound states, as the ground state at
least in the frustrated region.  Furthermore, the $XY$-type anisotropy of the
leg interactions is expected to stabilize the nematic TLL state.

In this paper, we use the term ``nematic TLL" for the TLL characterized not
only by the formation of two-magnon bound pairs but also by the dominant
nematic four-spin correlation function.  In the nematic TLL state composed of
the bosons of two-magnon bound pairs,\cite{schulz,hikihara-etal-1} the nematic
four-spin and longitudinal two-spin correlations are dual to each other, and
the decay exponent of the former, $\eta^{++--}$, and that of the latter,
$\eta_\ell^{zz}$, obey the relation $\eta^{++--}\eta_\ell^{zz}=1$. [See
Eqs.~(\ref{eq:C_nem}) and (\ref{eq:C_spin_zz}) for the definitions of
$\eta^{++--}$ and $\eta_\ell^{zz}$, respectively.]  If
\hbox{$\eta^{++--}\!<\!1$}, the
nematic correlation is dominant, leading to the nematic TLL state
defined above.  On the other hand, the TLL state with the dominant longitudinal
two-spin correlation is not allowed in the present system, since, if
$\eta_\ell^{zz}<1$, the Umklapp scattering becomes relevant and the
TLL state turns into the N{\'e}el state.  The situation is distinct from that
of zigzag chains under a magnetic field, in which the TLL state accompanied
with two-magnon bound pairs and dominant incommensurate spin-density-wave
correlation appears in wide regions of the magnetic phase
diagrams.\cite{hikihara-etal-1,sudan-etal,OkunishiT2003,HikiharaMFK2010}

We note, at this junction, that ground states of the ladder system governed
by the Hamiltonian ${\cal H}$ have already been discussed in a few other
cases.~\cite{tsukano-takahashi,ladder-diffleg-tone,sekiguchi-hida}
We~\cite{ladder-diffleg-tone} have determined the ground-state phase diagram on
the $1/\Gamma_{\rm r}$ (\hbox{$\Gamma_{\rm r}\!\geq\!1.0$}) versus
$J_{{\rm l},b}$ plane in the case where \hbox{$J_{{\rm l},a}\!=\!\pm0.2$},
\hbox{$J_{{\rm r}}\Gamma_{\rm r}\!=\!-1.0$}, and
\hbox{$\Delta_{\rm l}\!=\!1.0$}.  The obtained phase diagrams consist of the
$XY$1 phase, the triplet-dimer phase, the partial ferrimagnetic phase (which is
called the non-collinear ferrimagnetic phase in the paper), and the Haldane
phase.  Interestingly, the direct-product triplet-dimer state becomes to be
the exact ground state~\cite{tsukano-takahashi, hikihara-etal-2} when
\hbox{$J_{{\rm l},b}=-J_{{\rm l},a}$} and \hbox{$\Gamma_{\rm r}\!\gsim\!1.2$}.
On the other hand, extending Tsukano and
Takahashi's work~\cite{tsukano-takahashi}, Sekiguchi and
Hida~\cite{sekiguchi-hida} have shown, in the case where
\hbox{$J_{{\rm l},a} J_{{\rm l},b}\!<\!0$} and
\hbox{$\Delta_{\rm l}\!=\!\Gamma_{\rm r}\!=\!1.0$}, that the partial
ferrimagnetic state, which is a spontaneously magnetized TLL state with
incommensurate magnetic correlation, appears as the ground state over a
wide parameter range.

The remainder of this paper is organized as follows.  In Sect.~2, with the
help of physical considerations we numerically determine{,
by using the exact diagonalization (ED) method,}
the ground-state phase
diagram on the $\Delta_{\rm l}$ versus $J_{{\rm l},b}$ plane in the case where
\hbox{$J_{{\rm l},a}\!=\!0.2$}, \hbox{$J_{{\rm r}}\!=\!-1.0$}, and
\hbox{$\Gamma_{\rm r}\!=\!0.5$}.  In Sect.~3, we show that the distinguishing
features of the obtained phase diagram are well explained by performing
perturbational calculations.  In Sect.~4, in order to supplement the
reliability of the phase diagram, we carry out density-matrix
renormalization-group (DMRG) calculations~\cite{dmrg-white-1,dmrg-white-2}
for some physical quantities such as
the energy gaps, the local magnetization, and the spin correlation functions.
The final section (Sect.~5) contains concluding remarks.  In Appendix, the
ground-state phase diagram on the $\Gamma_{\rm r}$ versus $J_{{\rm l},b}$ plane
in the case where \hbox{$J_{{\rm l},a}\!=\!0.2$},
\hbox{$J_{{\rm r}}\!=\!-1.0$}, and \hbox{$\Delta_{\rm l}\!=\!1.0$} is briefly
discussed.

\section{Numerical Determination of the Ground-State Phase Diagram}

\begin{figure}[t]
   \begin{center}
       \scalebox{0.45}{\includegraphics{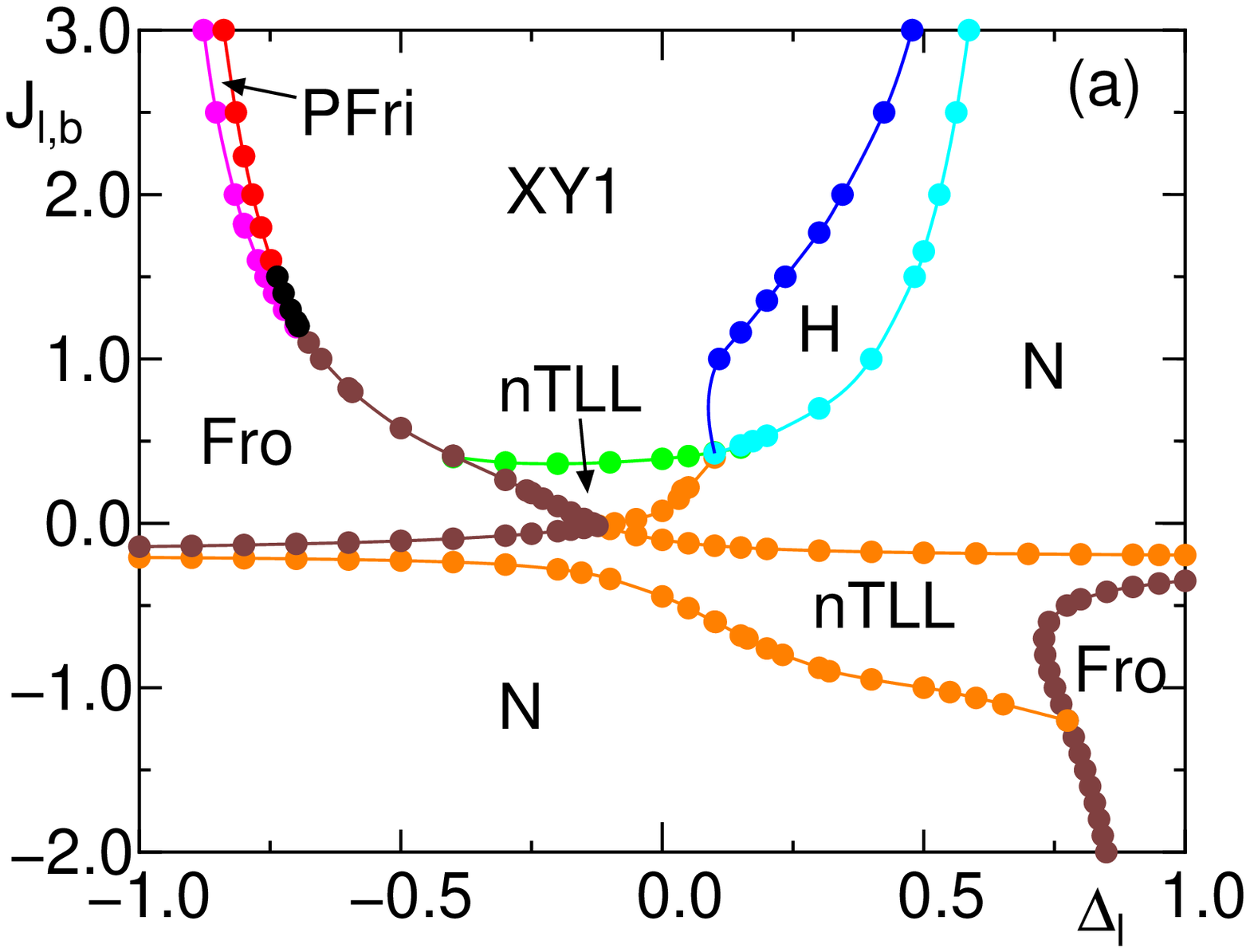}}
   \end{center}
   \begin{center}
       \scalebox{0.45}{\includegraphics{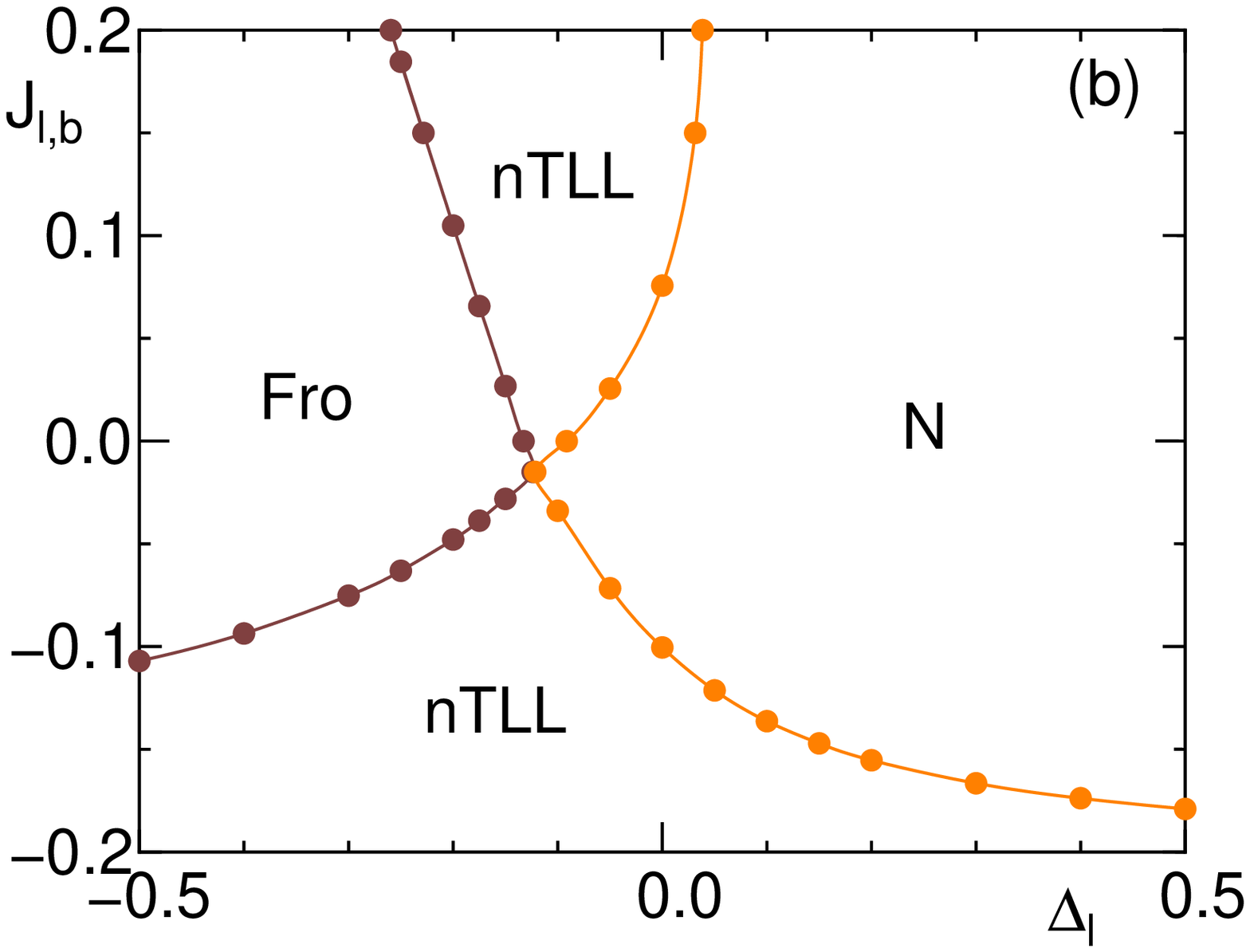}}~~~~
   \end{center}
   \caption{(Color online)
    Ground-state phase diagram on the $\Delta_{\rm l}$ versus
    $J_{{\rm l},b}$ plane in the case where \hbox{$J_{{\rm l},a}\!=\!0.2$},
    \hbox{$J_{{\rm r}}\!=\!-1.0$}, and \hbox{$\Gamma_{\rm r}\!=\!0.5$}, 
    obtained in the present work.  In (a) the whole view is shown, while
    in (b) an enlarged view around the origin is given.  The regions
    designated by Fro, PFri, XY1, H, N, and nTLL are those of the 
    ferromagnetic, partial ferrimagnetic, $XY1$, Haldane, N{\'e}el, and
    nematic TLL phases, respectively.}
   \label{fig:phasediagram}
\end{figure}

In this section we present the result for the ground-state phase diagram on
the $\Delta_{\rm l}$ versus $J_{{\rm l},b}$ plane in the case where
\hbox{$J_{{\rm l},a}\!=\!0.2$}, \hbox{$J_{{\rm r}}\!=\!-1.0$}, and
\hbox{$\Gamma_{\rm r}\!=\!0.5$}, which has been determined by using a variety
of numerical methods based on the {ED} calculation.
First of all, we show the obtained phase diagram in
Fig.~\ref{fig:phasediagram}. This phase diagram consists of six kinds of
phases; these are the
ferromagnetic, partial ferrimagnetic, $XY1$, Haldane, N{\'e}el, and nematic TLL
phases. It is noted that the nematic TLL phase appears as the ground state in the
strong-rung frustrated region (\hbox{$-1.0\lsim J_{{\rm l},b}\!<\!0.0$}), as
is expected, and survives even in the strong-rung unfrustrated
region (\hbox{$0.0\!\leq\!J_{{\rm l},b}\lsim0.4$}).

Before explaining how to numerically determine the phase boundary lines,
we introduce some related physical quantities.  We denote, respectively, by
$E_0^{\rm P}(L,M)$ and $E_1^{\rm P}(L,M)$, the lowest- and 
second-lowest-energy eigenvalues of the Hamiltonian ${\cal H}$ under the
periodic
boundary condition (PBC), \hbox{$\vec{S}_{L+1,\ell}\!=\!\vec{S}_{1,\ell}$},
within the subspace characterized by $L$ and the total magnetization
\hbox{$M\!\equiv\!\sum_{j=1}^{L} (S_{j,a}^z\!+\!S_{j,b}^z)$}.  Here, $M$ is
a good quantum number with the eigenvalues of \hbox{$M\!=\!0$}, $\pm 1$,
$\cdots$, $\pm L$.  Furthermore, the quantity $E_0^{\rm T}(L,M)$ is
the lowest-energy eigenvalue of ${\cal H}$  
under the twisted boundary condition (TBC),
\hbox{$S_{L+1,\ell}^x\!=\!-S_{1,\ell}^x$},
\hbox{$S_{L+1,\ell}^y\!=\!-S_{1,\ell}^y$}, and
\hbox{$S_{L+1,\ell}^z\!=\! S_{1,\ell}^z$}, within the subspace determined by
$L$ and $M$.  Then, we define several energy differences as follows:
\begin{eqnarray}
  && \!\!\!\!\!\!\!\!\!\!
  \Delta E_{0}^{\rm P}(L,M)=E_0^{\rm P}(L,M)-E_0^{\rm P}(L,0)\,, \\
  && \!\!\!\!\!\!\!\!\!\!
  \Delta E_{1}^{\rm P}(L,0)=E_1^{\rm P}(L,0)-E_0^{\rm P}(L,0)\,, \\
  && \!\!\!\!\!\!\!\!\!\!
  \Delta E_{0}^{\rm T}(L,0)=E_0^{\rm T}(L,0)-E_0^{\rm P}(L,0)\,.
\end{eqnarray}
It is noted that the values of the ground-state magnetization $M_{\rm g}(L)$
are \hbox{$M_{\rm g}(L)\!=\!L$}, \hbox{$0\!<\!M_{\rm g}(L)\!<\!L$},
and \hbox{$M_{\rm g}(L)\!=\!0$} in the ferromagnetic, partial ferrimagnetic,
and remaining four phases, respectively.

\begin{figure}[t]
   \begin{center}
       \scalebox{0.31}{\includegraphics{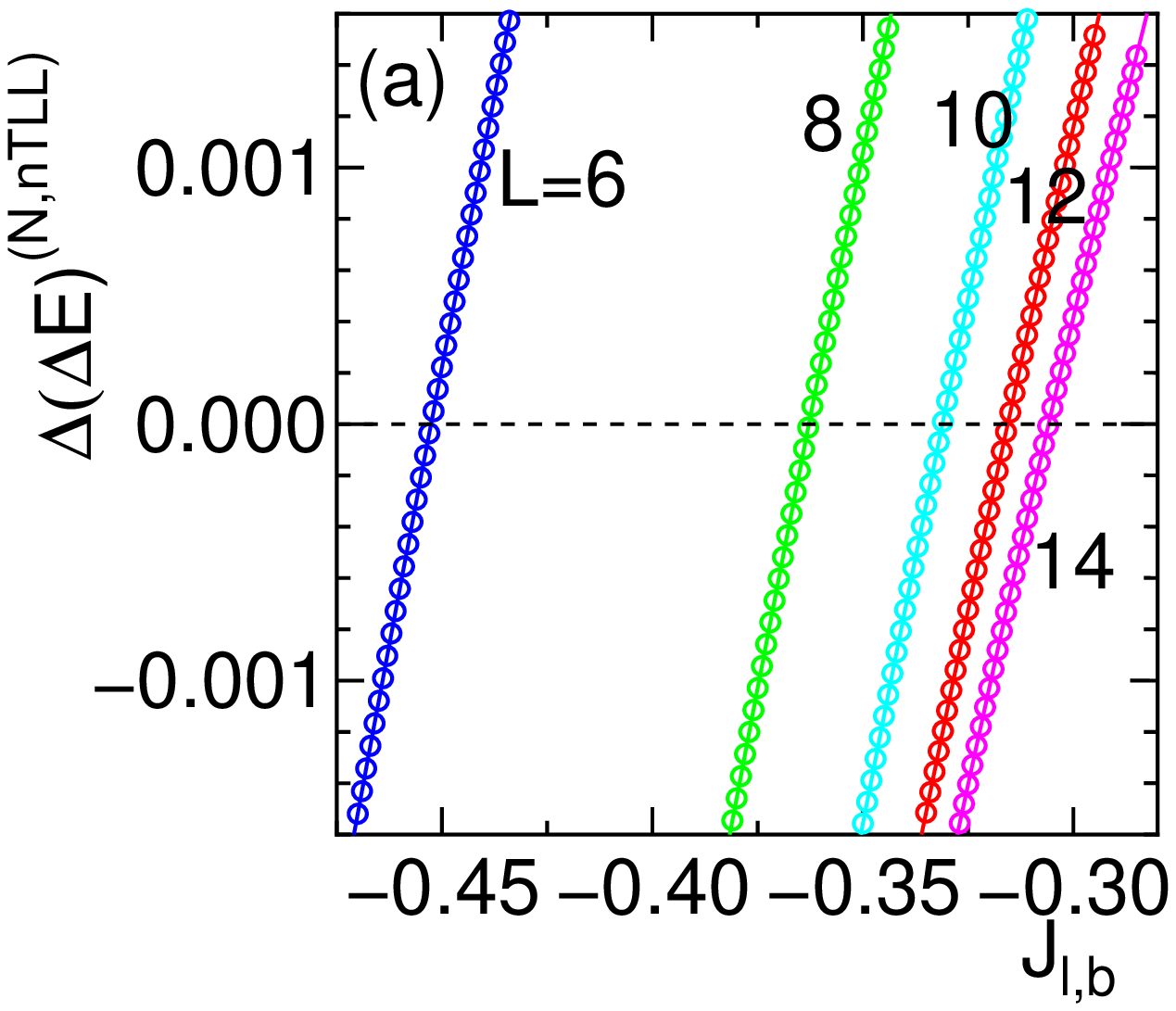}}~~
       \scalebox{0.31}{\includegraphics{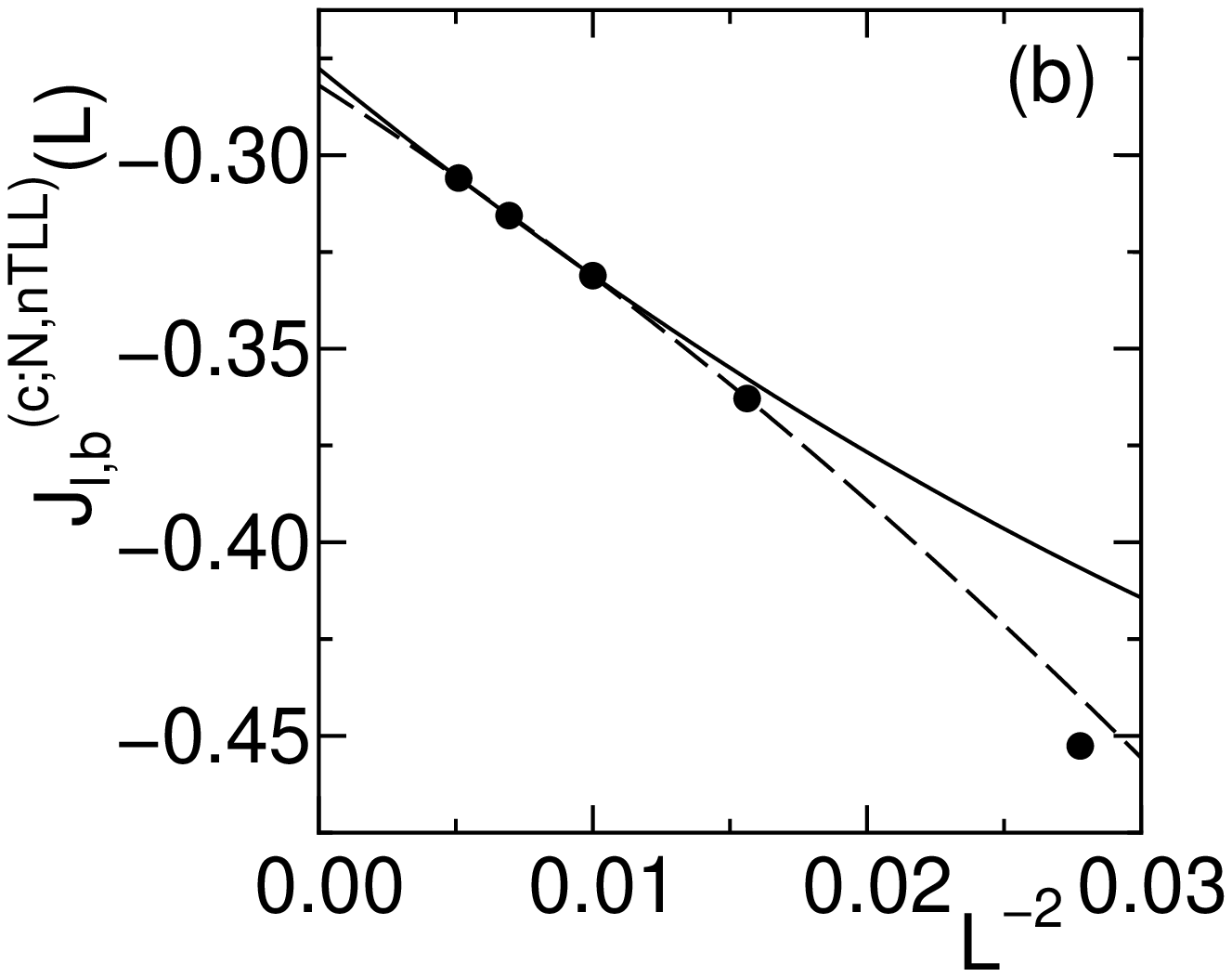}}
   \end{center}
   \caption{(Color online) (a) Plot of 
   $\Delta(\Delta E)^{\rm{(N,nTLL)}}
   [\equiv\!\Delta E_{1}^{\rm P}(L,0)\!-\!\frac{1}{2}
   \Delta E_{0}^{\rm P}(L,2)$] versus $J_{{\rm l},b}$ for
   \hbox{$\Delta_{\rm l}\!=\!-0.2$} and for \hbox{$L\!=\!6$}
   (blue), \hbox{$L\!=\!8$} (green), \hbox{$L\!=\!10$} (cyan),
   \hbox{$L\!=\!12$} (red), and \hbox{$L\!=\!14$} (magenta).
   The finite-size critical value $J_{{\rm l},b}^{({\rm c;N,nTLL})}(L)$
   for each $L$ is the value of $J_{{\rm l},b}$ at which
   \hbox{$\Delta(\Delta E)^{\rm{(N,nTLL)}}\!=\!0$} is satisfied.
   (b) Plot of $J_{{\rm l},b}^{({\rm c;N,nTLL})}(L)$ versus $L^{-2}$ for
   \hbox{$\Delta_{\rm l}\!=\!-0.2$}.  The solid and dashed lines represent,
   respectively, the least-squares fittings where the data from
   \hbox{$L\!=\!10$} to \hbox{$L\!=\!14$} and the data from \hbox{$L\!=\!8$}
   to \hbox{$L\!=\!14$} are used.  These extrapolations give the result
   \hbox{$J_{{\rm l},b}^{({\rm c;N,nTLL})}\!=\!-0.280\pm 0.002$} at the
   \hbox{$L\!\to\!\infty$} limit, where the numerical error is estimated
   from the difference between the above two extrapolated results.
   }
   \label{fig:N-nTLL}
\end{figure}

Let us first discuss the phase boundary lines between the N{\'e}el and nematic
TLL phases, which are depicted by the brown lines in
Fig.~\ref{fig:phasediagram}.  It should be emphasized that the nematic TLL
state is the TLL state which accompanies two-magnon bound states and is
equivalent to the $XY$2 state, originally proposed by Schulz.~\cite{schulz}
The corresponding phase transition is of the Berezinskii-Kosterlitz-Thouless
(BKT) type~\cite{BKT-1,BKT-2} accompanying the spontaneous
translational-symmetry breaking (STSB), and therefore, as is well known, the
phase boundary line can be accurately estimated by the level spectroscopy (LS)
method developed by Okamoto and Nomura.~\cite{okamoto-nomura,nomura-okamoto-1,
nomura-okamoto-2}  In this method, the
difficulty coming from the logarithmic-correction problem associated with
the BKT transition is removed.  According to this method, the critical value
$J_{{\rm l},b}^{({\rm c;N,nTLL})}$ of $J_{{\rm l},b}$ for a given value of
$\Delta_{\rm l}$ can be evaluated as follows.  We numerically solve the
equation
\begin{equation}
    \Delta E_{1}^{\rm P}(L,0) = \frac{1}{2}\,\Delta E_{0}^{\rm P}(L,2)\,,
    \label{eq:N-nTLL}
\end{equation}
to compute the finite-size critical value
$J_{{\rm l},b}^{({\rm c;N,nTLL})}(L)$ for various values of $L$.   Then,
we extrapolate these finite-size results to the thermodynamic
(\hbox{$L\!\to\!\infty$}) limit.  Practically, we have carried out the ED
calculation to compute $J_{{\rm l},b}^{({\rm c;N,nTLL})}(L)$ for finite-$L$
systems with \hbox{$L\!=\!6$}, $8$, $\cdots$, $14$ rungs.  Performing the
\hbox{$L\!\to\!\infty$} extrapolation of
$J_{{\rm l},b}^{({\rm c;N,nTLL})}(L)$'s, we have fitted them
to quadratic functions of $L^{-2}$ by using the least-squares method.  An
example of the results is
\hbox{$J_{{\rm l},b}^{({\rm c;N,nTLL})}\!=\!-0.280\pm 0.002$} for
\hbox{$\Delta_{\rm l}\!=\!-0.2$}.  These procedures for the evaluation of
$J_{{\rm l},b}^{({\rm c;N,nTLL})}$ are illustrated in Fig.~\ref{fig:N-nTLL}.
The phase boundary lines are drawn by plotting
$J_{{\rm l},b}^{({\rm c;N,nTLL})}$'s as functions of $\Delta_{\rm l}$.

In the following discussion in this section, we mainly describe how to
compute the finite-size critical values only, since the evaluation of the
critical value at the thermodynamic limit has been done in the same
way.

\begin{figure}[t]
   \begin{center}
       \scalebox{0.31}{\includegraphics{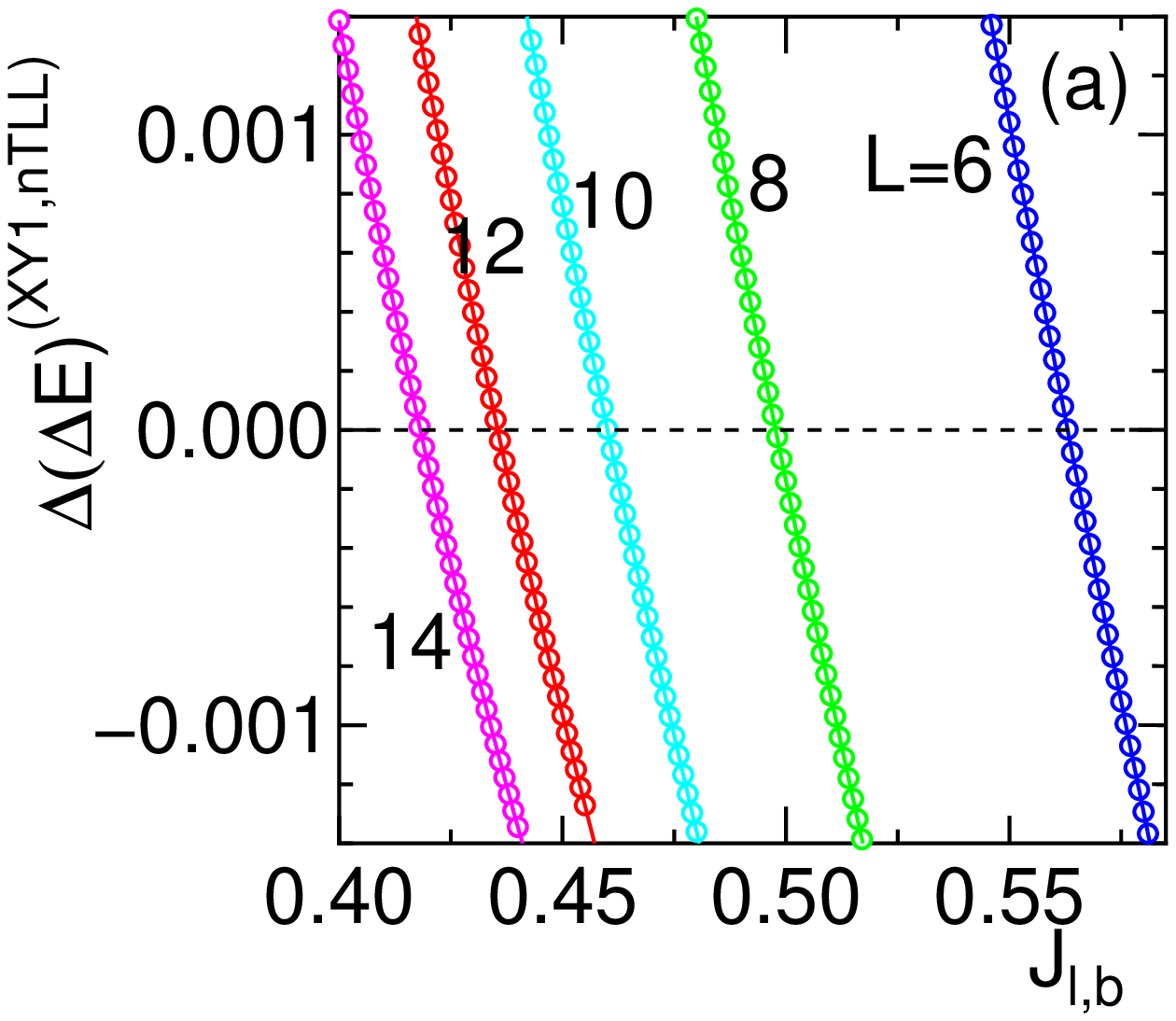}}~~
       \scalebox{0.31}{\includegraphics{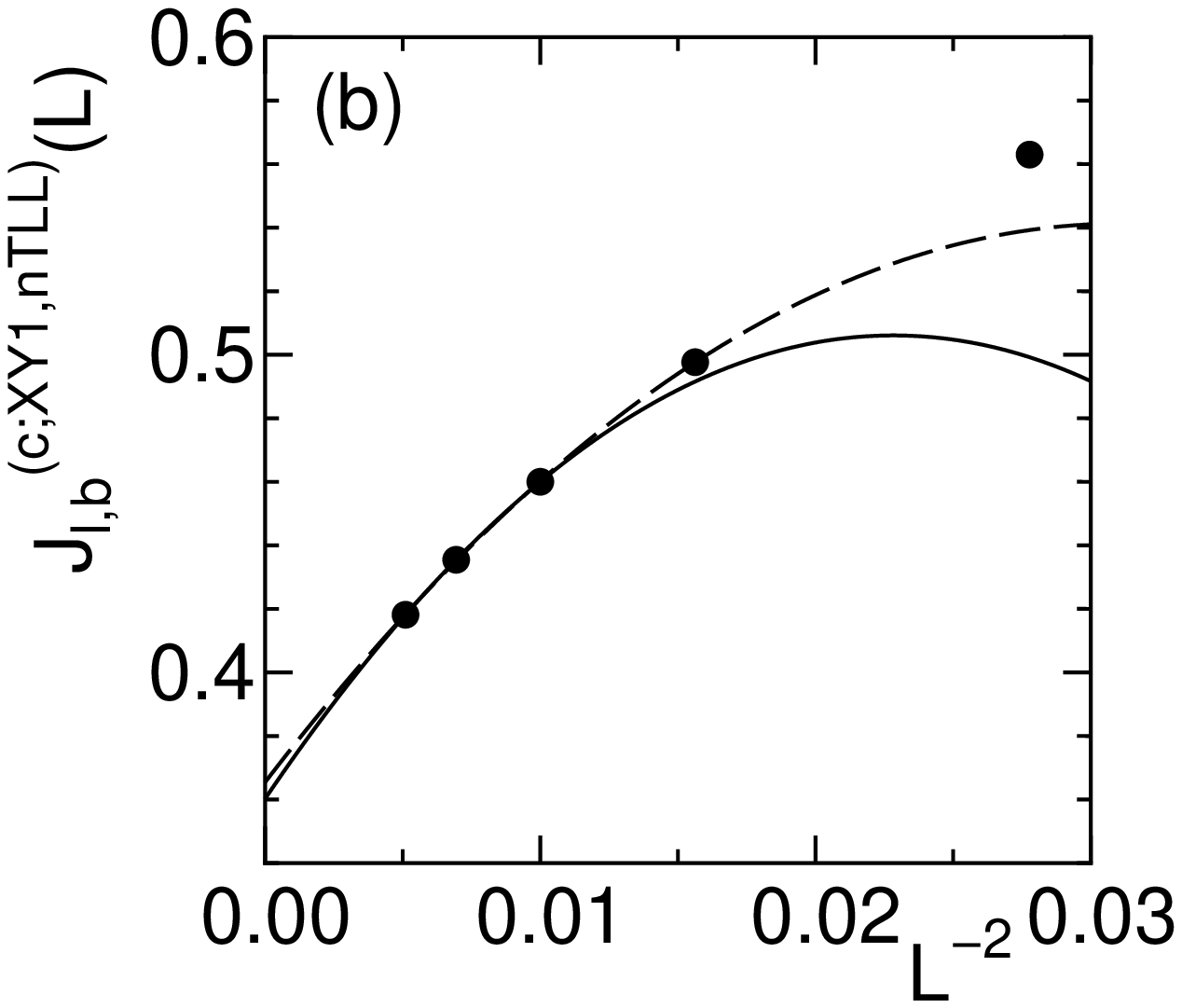}}
   \end{center}
   \caption{(Color online)
   (a) Plot of $\Delta(\Delta E)^{(XY{\rm 1,nTLL})}
   [\equiv\!\Delta E_{0}^{\rm P}(L,1)\!-\!\frac{1}{2}
   \Delta E_{0}^{\rm P}(L,2)$] versus $J_{{\rm l},b}$ for
   \hbox{$\Delta_{\rm l}\!=\!-0.2$}.  The correspondence of the color of
   each line to the value of each $L$ is the same as that in
   Fig.~\ref{fig:N-nTLL}(a).  (b) Plot of
   $J_{{\rm l},b}^{({\rm c;}XY{\rm1,nTLL})}(L)$ versus $L^{-2}$ for
   \hbox{$\Delta_{\rm l}\!=\!-0.2$}.  The meanings of the solid and dashed
   lines  are the same as those in Fig.~\ref{fig:N-nTLL}(b).  The extrapolated
   value to \hbox{$L\!\to\!\infty$} is given by 
   \hbox{$J_{{\rm l},b}^{({\rm c;}XY{\rm1,nTLL})}\!=\!0.363\pm 0.003$}.
   }
   \label{fig:XY1-nTLL}
\end{figure}

Secondly (see the green line in Fig.~\ref{fig:phasediagram}), we discuss
the phase boundary line between the $XY$1 and nematic TLL phases.  The nematic
TLL state accompanies two-magnon bound states, as mentioned above, while the
$XY$1 state does not.  These lead to the following fact.  In the ground-state
magnetization curve for a given finite-size system under an external magnetic
field ${\vec H}$ applied along the $z$-axis, the magnetization increases
stepwisely with increasing $\vert {\vec H}\vert$; the first step occurs from
the \hbox{$M_{\rm g}(L)\!=\!0$} state to the \hbox{$M_{\rm g}(L)\!=\!2$} state
in the nematic TLL phase, while it occurs from the \hbox{$M_{\rm g}(L)\!=\!0$}
state to the \hbox{$M_{\rm g}(L)\!=\!1$} state in the $XY$1 phase.  From these
arguments, it is easy to see that the finite-size critical value
$J_{{\rm l},b}^{({\rm c;}XY{\rm1,nTLL})}(L)$ of $J_{{\rm l},b}$, for a given
value of $\Delta_{\rm l}$, for the phase transition between the $XY$1 and
nematic TLL phases is calculated by solving
\begin{equation}
    \Delta E_{0}^{\rm P}(L,1) = \frac{1}{2}\,\Delta E_{0}^{\rm P}(L,2)
    \label{eq:XY1-nTLL}
\end{equation}
numerically.  Figure~\ref{fig:XY1-nTLL} shows the determination of the critical
value $J_{{\rm l},b}^{({\rm c;}XY{\rm1,nTLL})}$ at the thermodynamic limit for
\hbox{$\Delta_{\rm l}\!=\!-0.2$}.

\begin{figure}[t]
   \begin{center}
       \scalebox{0.31}{\includegraphics{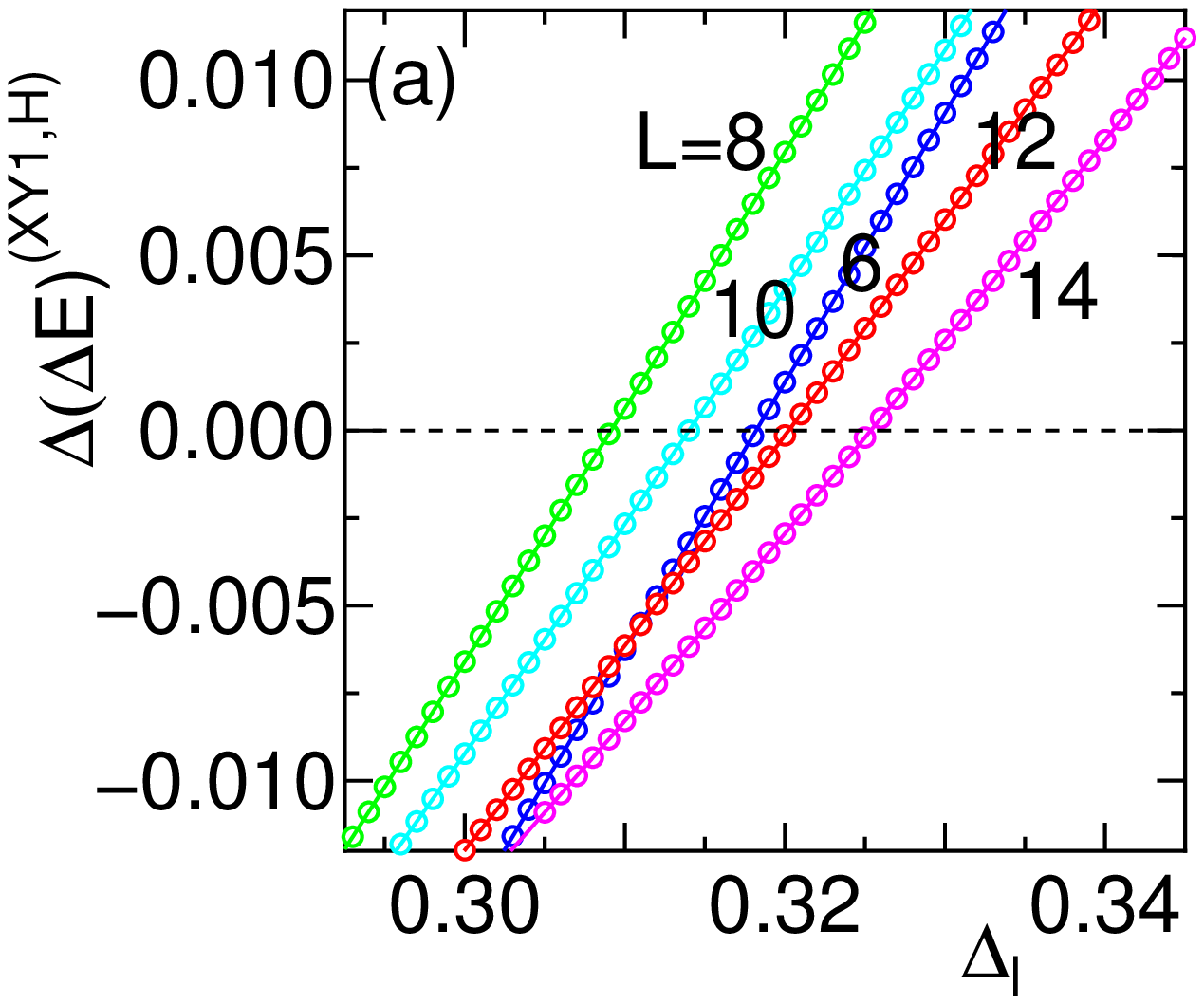}}~~
       \scalebox{0.31}{\includegraphics{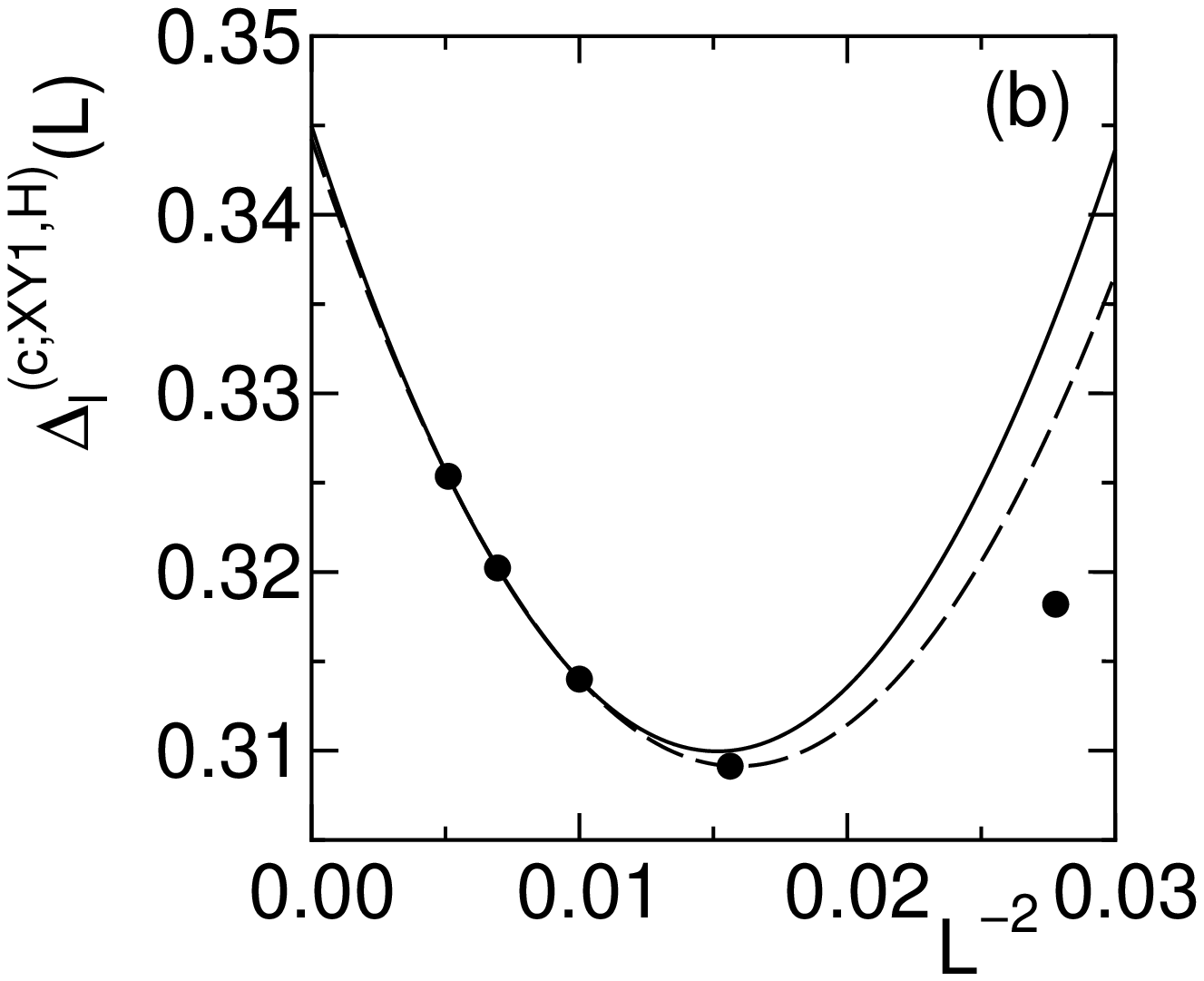}}
   \end{center}
   \caption{(Color online)
   (a) Plot of $\Delta(\Delta E)^{(XY{\rm 1,H})}
   [\equiv\!\Delta E_{0}^{\rm P}(L,2)\!-\!
   \Delta E_{0}^{\rm T}(L,0)$] versus $\Delta_{\rm l}$ for
   \hbox{$J_{{\rm l},b}\!=\!2.0$}.  The correspondence of the color of
   each line to the value of each $L$ is the same as that in
   Fig.~\ref{fig:N-nTLL}(a).  (b) Plot of
   $\Delta_{\rm l}^{({\rm c;}XY{\rm 1,H)}}(L)$ versus $L^{-2}$ for
   \hbox{$J_{{\rm l},b}\!=\!2.0$}.  The meanings of the solid and dashed
   lines  are the same as those in Fig.~\ref{fig:N-nTLL}(b).  The extrapolated
   value to \hbox{$L\!\to\!\infty$} is given by 
   \hbox{$\Delta_{\rm l}^{({\rm c;}XY{\rm 1,H)}}\!=\!0.345\pm 0.001$}.
   }
   \label{fig:XY1-H}
\end{figure}

Thirdly (the blue line in Fig.~\ref{fig:phasediagram}), the phase transition
between the $XY$1 and Haldane phases is the BKT transition without the STSB.
It is also well known that, in this case, Nomura and Kitazawa's LS
method,~\cite{LS-NK} in which the difficulty coming from the
logarithmic-correction problem is also removed, is very powerful for
determining the phase boundary line.  The finite-size critical value
$\Delta_{\rm l}^{({\rm c;}XY{\rm 1,H)}}(L)$ of $\Delta_{\rm l}$ for a given
value of $J_{{\rm l},b}$ is calculated by using the equation
\begin{equation}
    \Delta E_{0}^{\rm P}(L,2) = \Delta E_{0}^{\rm T}(L,0)\,.
    \label{eq:XY1-H}
\end{equation}
In Fig.~\ref{fig:XY1-H} we show the determination of the critical
value $\Delta_{\rm l}^{({\rm c;}XY{\rm 1,H)}}$ at the thermodynamic limit for
\hbox{$J_{{\rm l},b}\!=\!2.0$}.

\begin{figure}[t]
   \begin{center}
       \scalebox{0.31}{\includegraphics{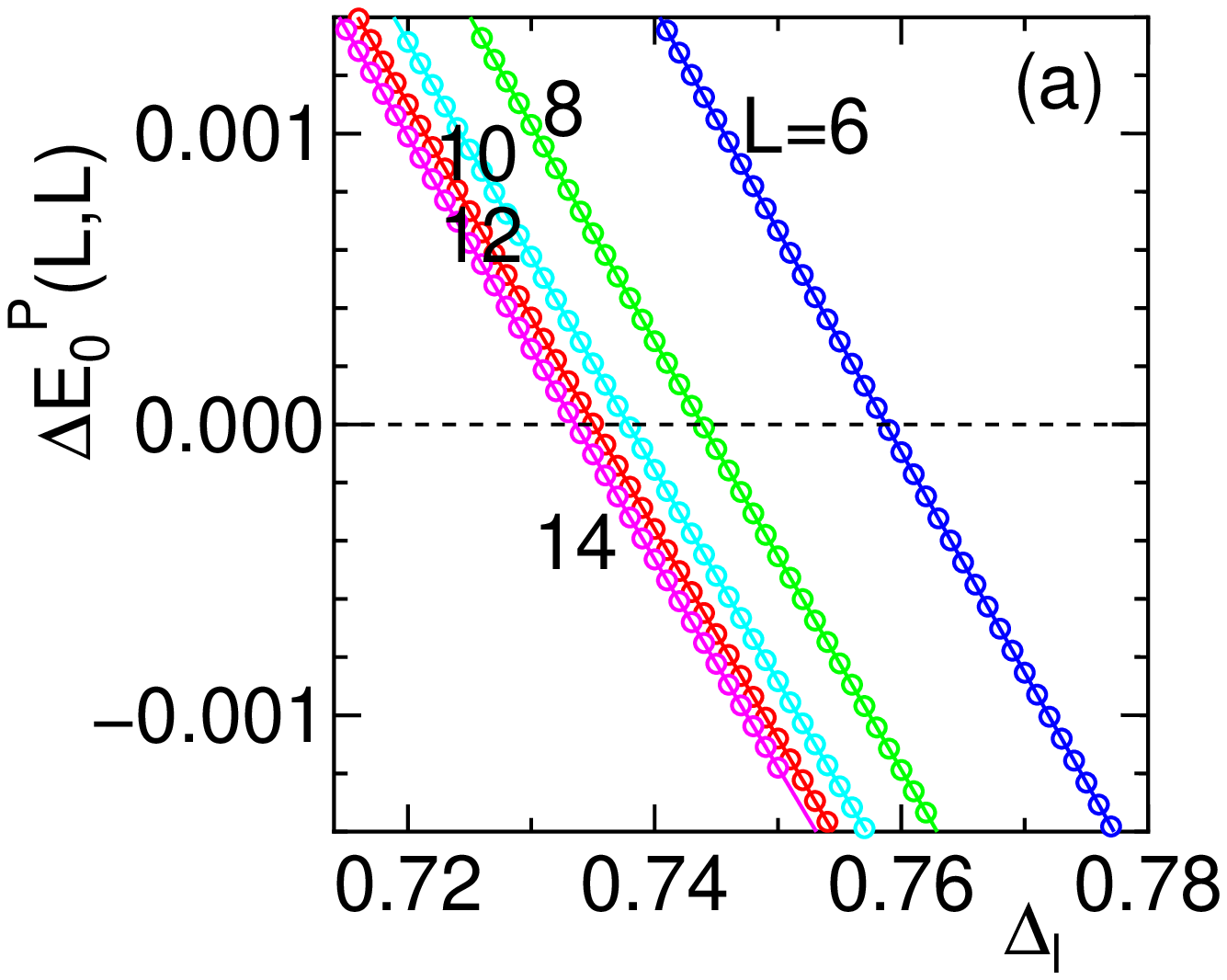}}~~
       \scalebox{0.31}{\includegraphics{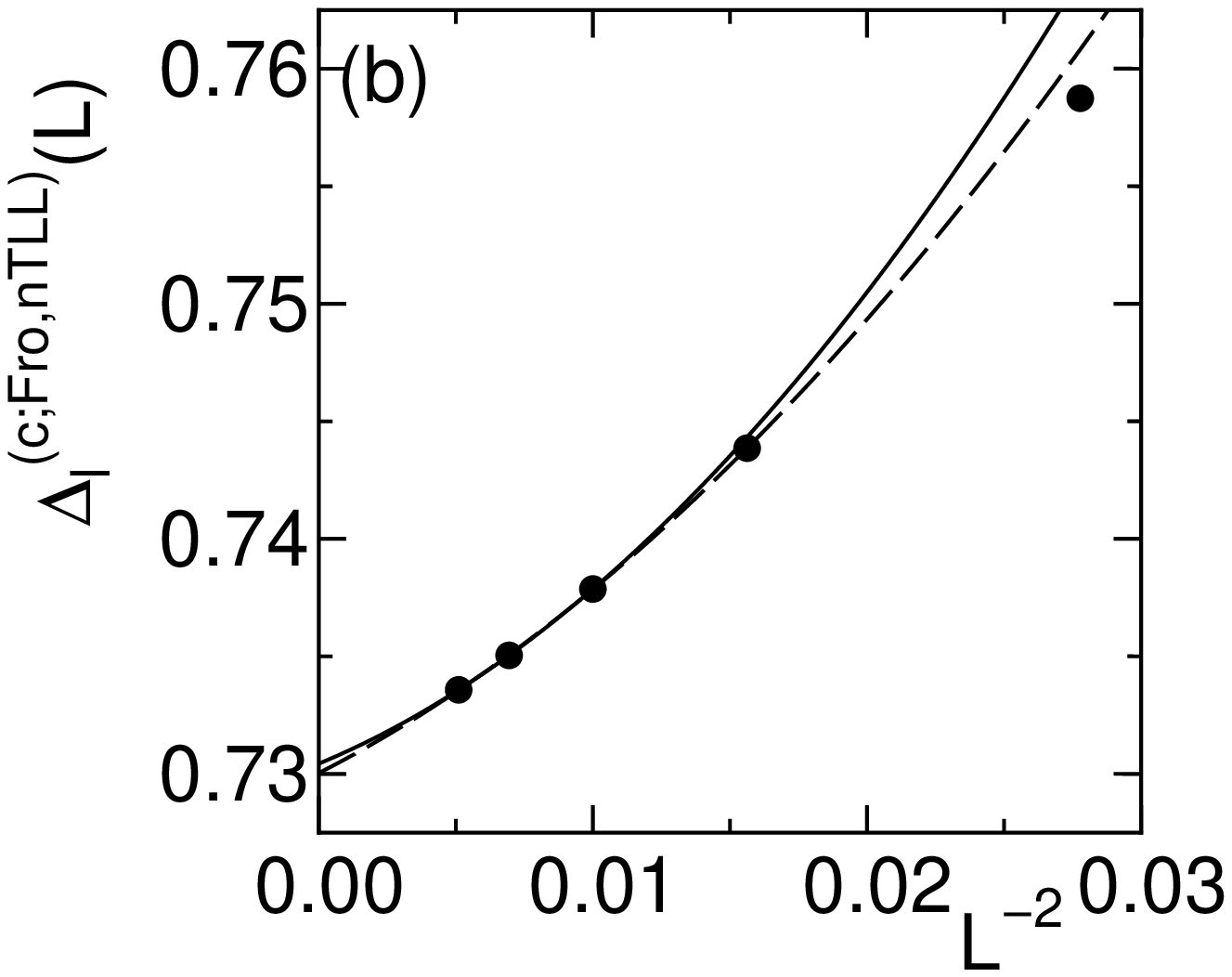}}
   \end{center}
   \caption{(Color online)
   (a) Plot of $\Delta E_{0}^{{\rm P}}(L,L)$
   versus $\Delta_{\rm l}$ for
   \hbox{$J_{{\rm l},b}\!=\!-0.7$}.  The correspondence of the color of
   each line to the value of each $L$ is the same as that in
   Fig.~\ref{fig:N-nTLL}(a).  (b) Plot of
   $\Delta_{\rm l}^{({\rm c;Fro,nTLL})}(L)$ versus $L^{-2}$ for
   \hbox{$J_{{\rm l},b}\!=\!-0.7$}.  The meanings of the solid and dashed
   lines  are the same as those in Fig.~\ref{fig:N-nTLL}(b).  The extrapolated
   value to \hbox{$L\!\to\!\infty$} is given by 
   \hbox{$\Delta_{\rm l}^{({\rm c;Fro,nTLL)}}\!=\!0.730\pm 0.001$}.
   }
   \label{fig:Fro-nTLL}
\end{figure}

Fourthly (the {umber} lines in Fig.~\ref{fig:phasediagram}),
the phase
transitions between the ferromagnetic phase and one of the $XY$1, N{\'e}el, and
nematic TLL phases are of the first-order, because these
transitions are between the $M_{\rm g}(L)\!=\!L$ and $M_{\rm g}(L)\!=\!0$
states at the \hbox{$L\!\to\!\infty$} limit.  Thus, it is apparent that the
phase boundary lines are obtained from the equation
\begin{equation}
    \Delta E_{0}^{\rm P}(L,L) = 0\,.
    \label{eq:H-M=0}
\end{equation}
As an example, we discuss here the phase transition between the
ferromagnetic and nematic TLL phases in the case of
\hbox{$J_{{\rm l},b}\!=\!-0.7$}, denoting the finite-size critical value
by $\Delta_{\rm l}^{({\rm c;Fro,nTLL})}(L)$.  In Fig.~\ref{fig:Fro-nTLL}
the determination of the
critical value $\Delta_{\rm l}^{({\rm c;Fro,nTLL})}$ at the thermodynamic
limit in this case is illustrated.

\begin{figure}[t]
   \begin{center}
       \scalebox{0.31}{\includegraphics{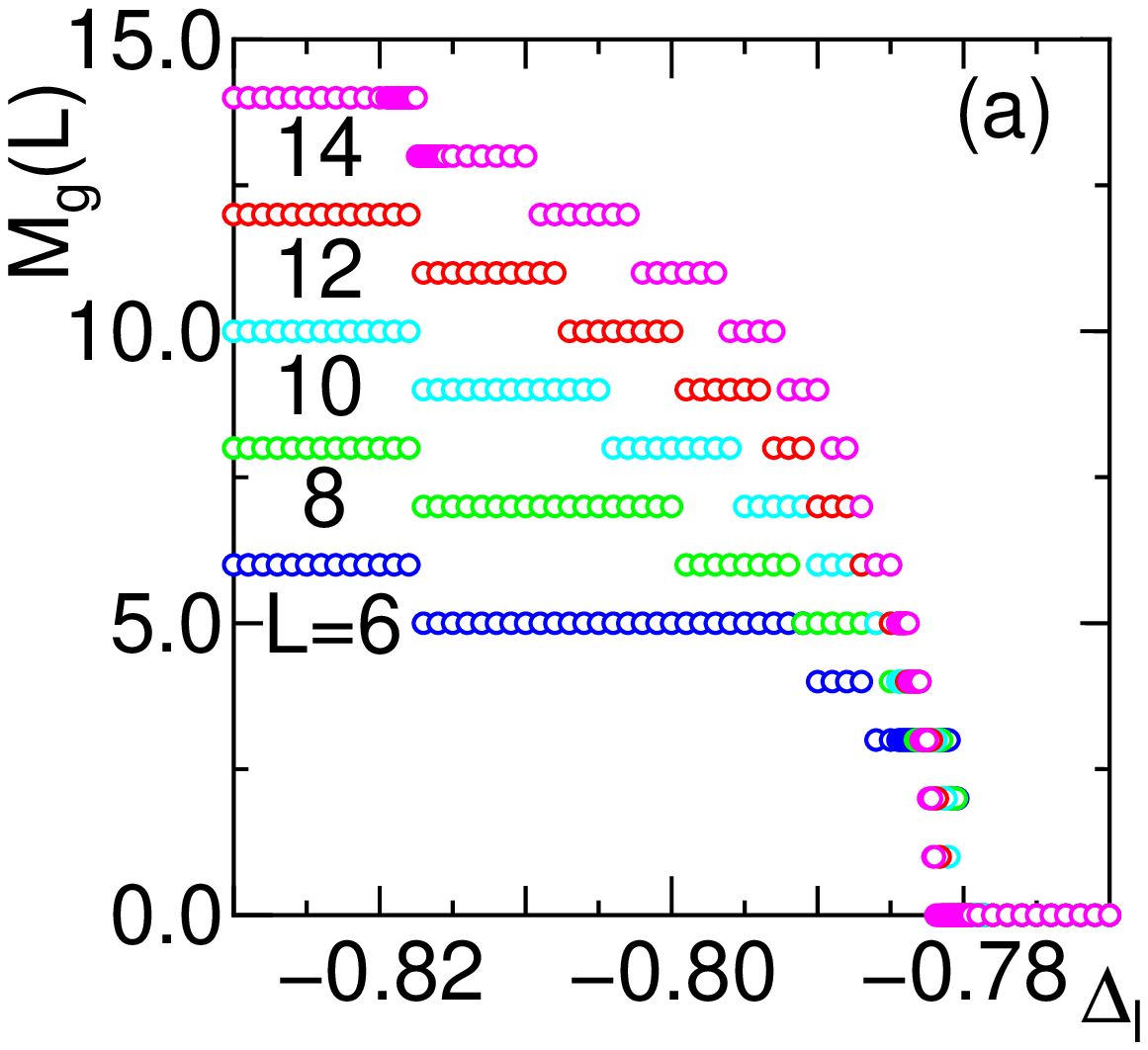}}~~
       \scalebox{0.31}{\includegraphics{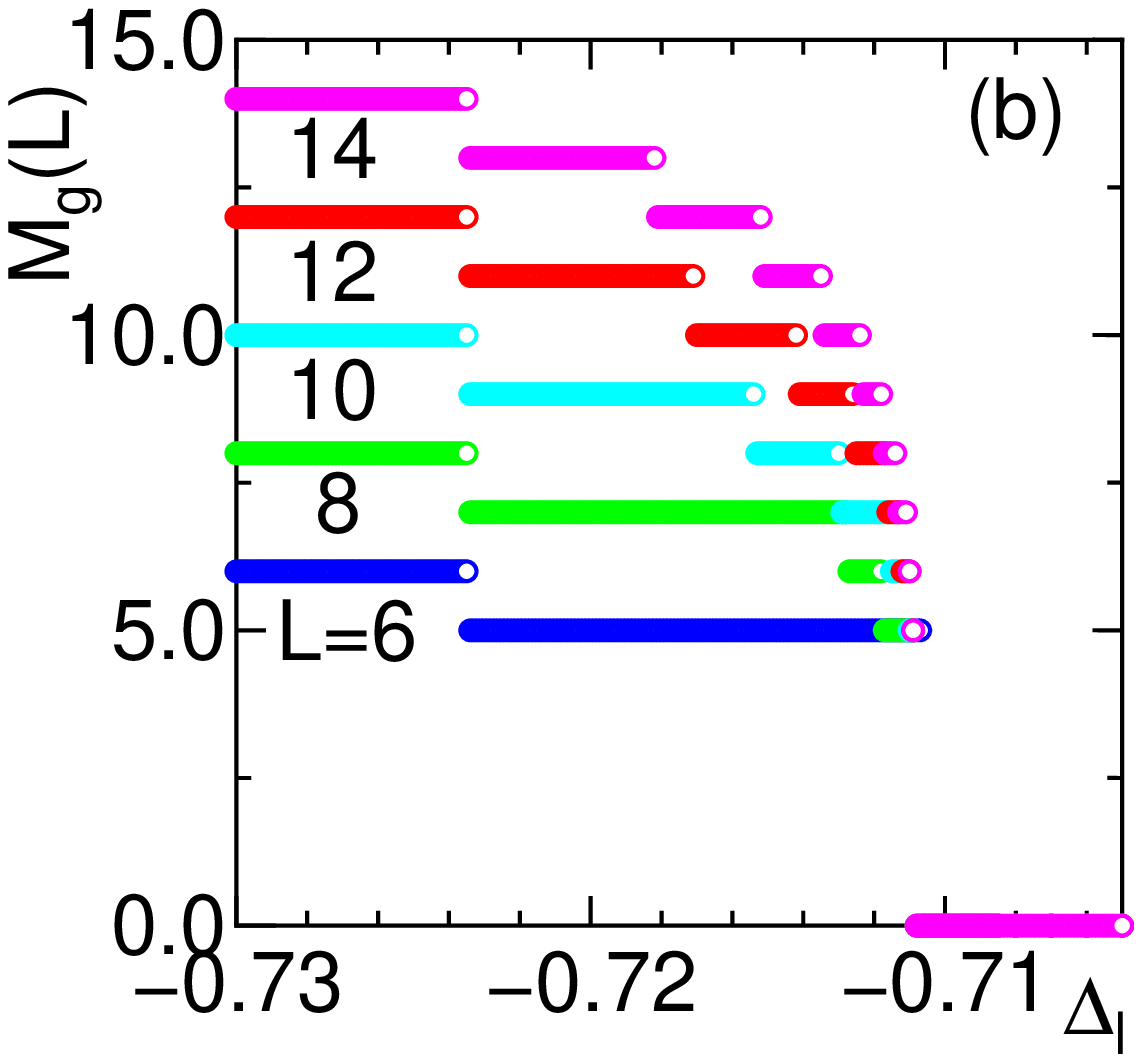}}
   \end{center}
   \caption{(Color online)
   Plot of $M_{\rm g}(L)$ versus $\Delta_{\rm l}$
   in the cases of (a) \hbox{$J_{{\rm l},b}\!=\!2.0$} and (b)
   \hbox{$J_{{\rm l},b}\!=\!1.3$}.  The correspondence of the color of
   each line to the value of each $L$ is the same as that in
   Fig.~\ref{fig:N-nTLL}(a).
   }
   \label{fig:grandmag}
\end{figure}

\begin{figure}[t]
   \begin{center}
       \scalebox{0.31}{\includegraphics{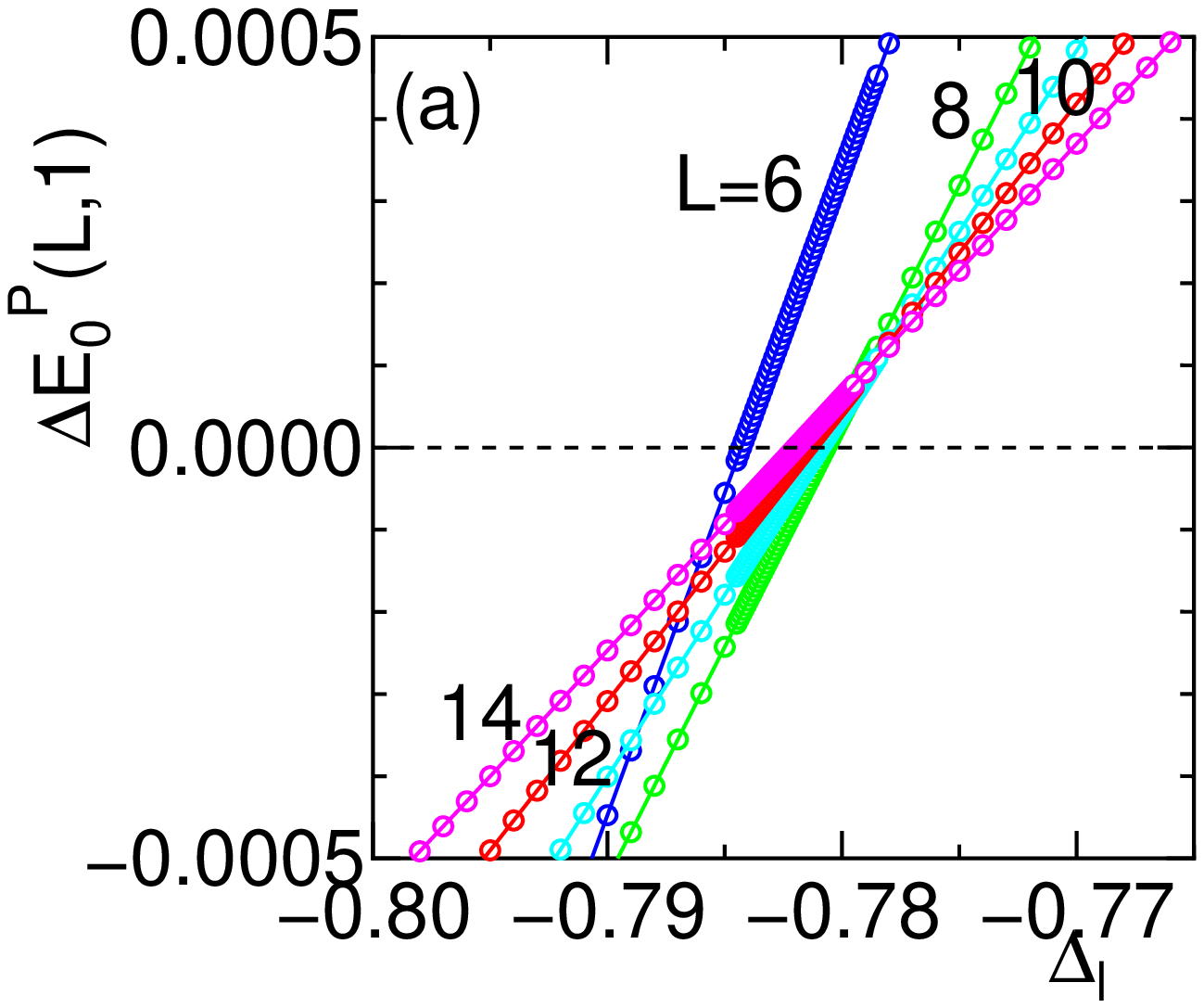}}~~
       \scalebox{0.31}{\includegraphics{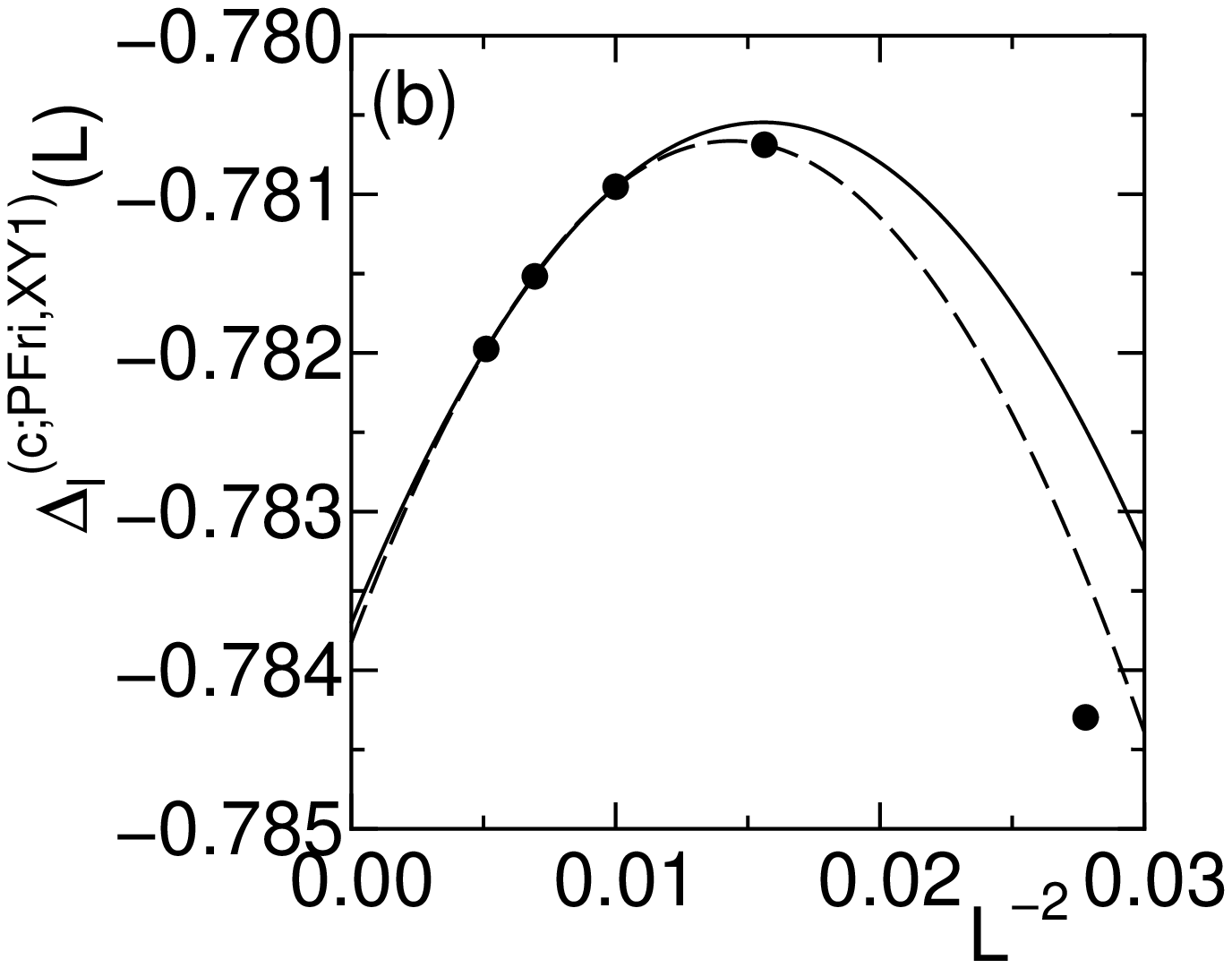}}
   \end{center}
   \caption{(Color online)
   (a) Plot of
   $\Delta E_{0}^{{\rm P}}(L,1)$ versus $\Delta_{\rm l}$ for
   \hbox{$J_{{\rm l},b}\!=\!2.0$}.  The correspondence of the color of
   each line to the value of each $L$ is the same as that in
   Fig.~\ref{fig:N-nTLL}(a).  (b) Plot of
   $\Delta_{\rm l}^{({\rm c;PFri,XY1})}(L)$ versus $L^{-2}$ for
   \hbox{$J_{{\rm l},b}\!=\!2.0$}.  The meanings of the solid and dashed
   lines are the same as those in Fig.~\ref{fig:N-nTLL}(b).  The extrapolated
   value to \hbox{$L\!\to\!\infty$} is given by 
   \hbox{$\Delta_{\rm l}^{({\rm c;PFri,XY1)}}\!=\!-0.784\pm 0.001$}.
   }
   \label{fig:PFri-XY1}
\end{figure}

Fifthly (the magenta, red, and black lines in Fig.~\ref{fig:phasediagram}),
according to the result of our calculations, the partial ferrimagnetic
phase appears in a rather narrow region of $\Delta_{\rm l}$, which lies between
the ferromagnetic-phase region and the $XY$1-phase region, only when
\hbox{$J_{{\rm l},b}\!\gsim\!1.2$}.  It shows further that the phase transition
between the ferromagnetic and partial ferrimagnetic phases is between the
$M_{\rm g}(L)\!=\!L$ and $M_{\rm g}(L)\!=\!L\!-\!1$ states, while that between
the partial ferrimagnetic and $XY$1 phases is between the
$M_{\rm g}(L)\!=\!1$ and $M_{\rm g}(L)\!=\!0$ states or between
the $L\!-\!1\!\geq\!M_{\rm g}(L)\!\geq\!2$ and $M_{\rm g}(L)\!=\!0$ states
depending upon whether \hbox{$J_{{\rm l},b}\!\gsim\!1.55$} or
\hbox{1.55\!\gsim\!$J_{{\rm l},b}\!\gsim\!1.2$}.  Thus, it may be considered
that the former transition is always of the second-order, while the latter
transition is of the second-order when \hbox{$J_{{\rm l},b}\!\gsim\!1.55$} and
of the first-order when \hbox{1.55\!\gsim\!$J_{{\rm l},b}\!\gsim\!1.2$}.
{The $\Delta_{\rm l}$-dependences of $M_{\rm g}(L)$'s for
\hbox{$L\!=\!6$}, $8$, $\cdots$, $14$} in the
cases of \hbox{$J_{{\rm l},b}\!=\!2.0$} and $1.3$ are depicted in
Figs.~\ref{fig:grandmag}(a) and (b), respectively.  The phase boundary line
between the ferromagnetic and partial ferrimagnetic phases can be
analytically calculated by examining the instability of the one-magnon
excitation from the ferromagnetic state.
However, since this calculation is straightforward but too tedious, we show
in Fig.~\ref{fig:phasediagram} the numerical result obtained by solving
\begin{equation}
    \Delta E_{0}^{\rm P}(L,L) = \Delta E_{0}^{\rm P}(L,L-1)\,.
    \label{eq:ferroinstability}
\end{equation}
It is noted that the obtained finite-size critical value is
independent of $L$
(the magenta line).  
This means that in the
present case, the $M_{\rm g}(L)=L-1$ state with the lowest energy
is a commensurate state.
The finite-size critical value
$\Delta_{\rm l}^{({\rm c;PFri,}XY{\rm 1})}(L)$ of $\Delta_{\rm l}$, for a given
value of $J_{{\rm l},b}$, for the second-order phase transition
between the partial ferrimagnetic and $XY$1 phases is calculated by solving
\begin{equation}
    \Delta E_{0}^{\rm P}(L,1) = 0\,.
    \label{eq:PFri-XY1}
\end{equation}
Figure~\ref{fig:PFri-XY1} illustrates the determination of the corresponding
critical value $\Delta_{\rm l}^{({\rm c;PFri,}XY{\rm 1})}$ at the
thermodynamic limit in the case where \hbox{$J_{{\rm l},b}\!=\!2.0$} (the red
line).  Generally speaking, it is difficult to accurately determine the
critical value $\Delta_{\rm l}^{({\rm c;PFri,}XY{\rm 1})}$ of
$\Delta_{\rm l}$ at the thermodynamic limit, for a given value of
$J_{{\rm l},b}$, for the first-order phase transition between the partial
ferrimagnetic and $XY$1 phases by performing only the ED calculations for
finite-$L$ systems with up to \hbox{$L\!=\!14$} rungs.  Fortunately, however,
in the present case the corresponding finite-size
critical value $\Delta_{\rm l}^{({\rm c;PFri,}XY{\rm 1})}(L)$ of
$\Delta_{\rm l}$ is almost independent of $L$, 
{as can been seen from Fig.~\ref{fig:grandmag}(b),}
and therefore we use
$\Delta_{\rm l}^{({\rm c;PFri,}XY{\rm 1})}(14)$ for
$\Delta_{\rm l}^{({\rm c;PFri,}XY{\rm 1})}$ (the black line).  We consider that
the numerical error due to this simplification is less than $0.001$.

\begin{figure}[t]
   \begin{center}
       \scalebox{0.31}{\includegraphics{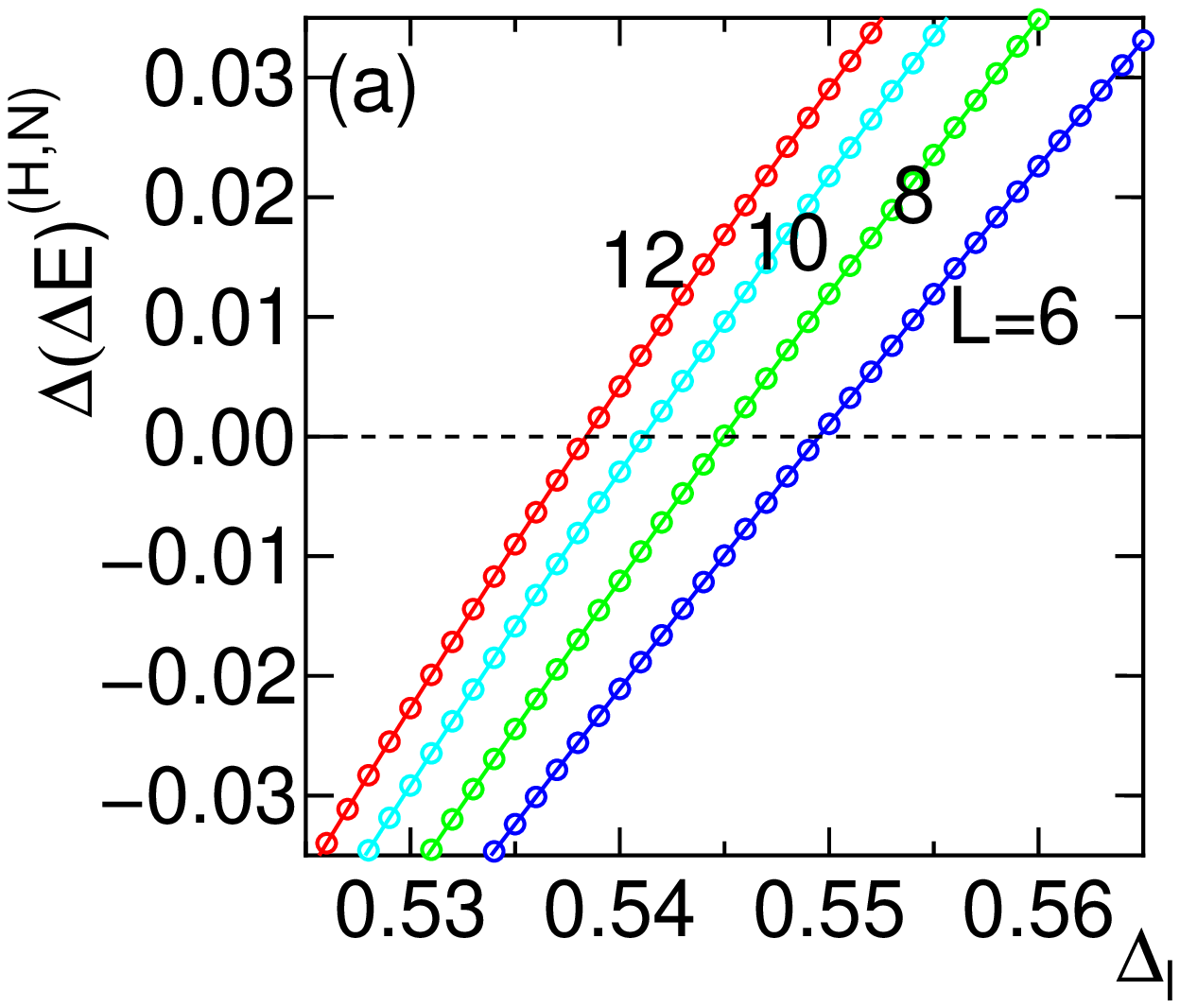}}~~
       \scalebox{0.31}{\includegraphics{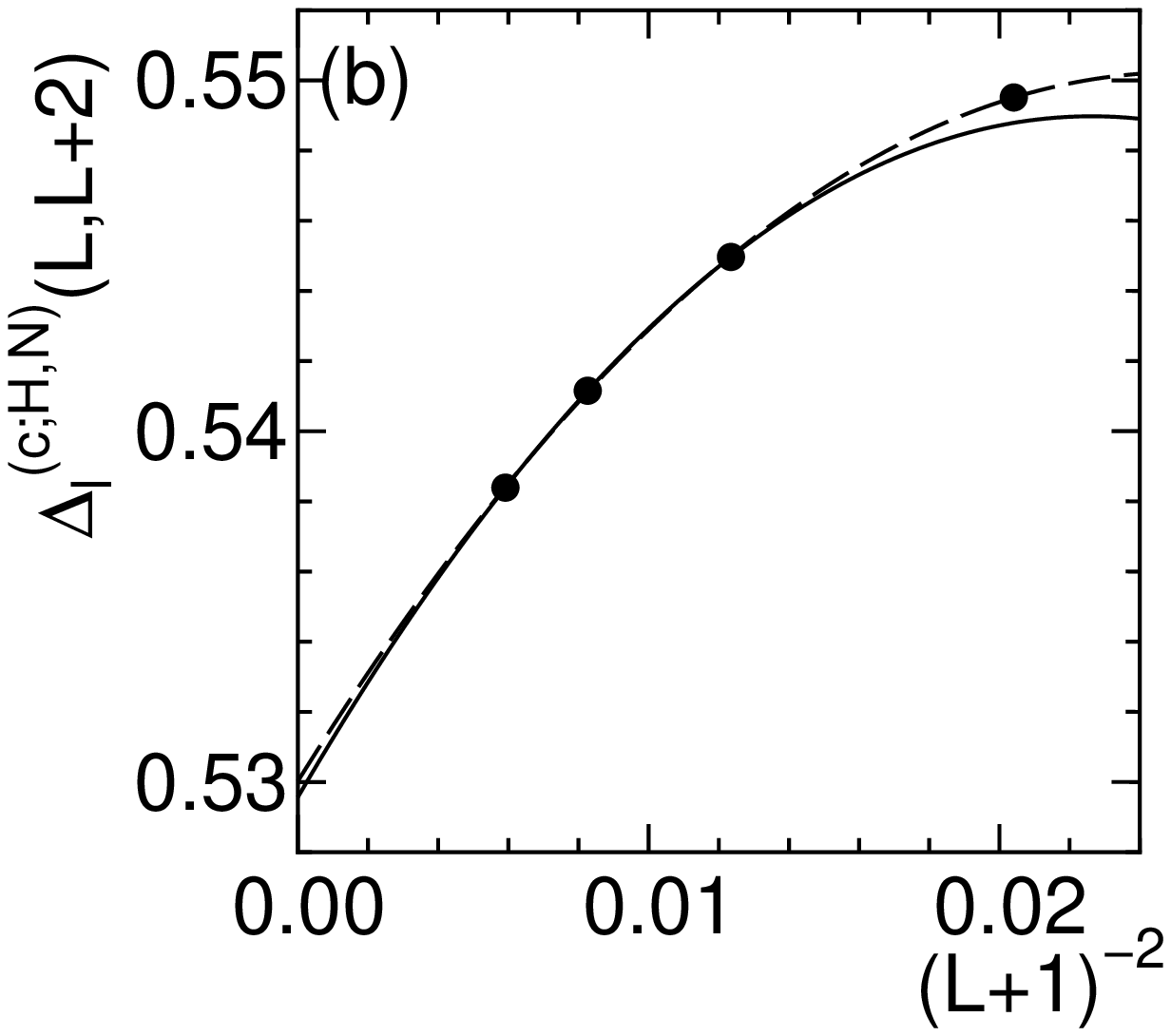}}
   \end{center}
   \caption{(Color online)
   (a) Plot of $\Delta(\Delta E)^{\rm{(H,N)}}
   [\equiv\!L\,\Delta E_{1}^{\rm P}(L,0)\!-\!
   (L+2)\,\Delta E_{1}^{\rm P}(L+2,0)$] versus $\Delta_{\rm l}$ for
   \hbox{$J_{{\rm l},b}\!=\!2.0$}.  The correspondence of the color of
   each line to the value of each $L$ is the same as that in
   Fig.~\ref{fig:N-nTLL}(a).  (b) Plot of
   $\Delta_{\rm l}^{({\rm c;H,N})}(L,L\!+\!2)$ versus $(L+1)^{-2}$ for
   \hbox{$J_{{\rm l},b}\!=\!2.0$}.  The solid and dashed lines represent,
   respectively, the least-squares fittings where the data from
   \hbox{$L\!=\!8$} to \hbox{$L\!=\!12$} and the data from \hbox{$L\!=\!6$}
   to \hbox{$L\!=\!12$} are used.  The extrapolated value to 
   \hbox{$L\!\to\!\infty$} is given by 
   \hbox{$\Delta_{\rm l}^{({\rm c;H,N})}\!=\!0.530\pm 0.001$}.
   }
   \label{fig:H-N}
\end{figure}

Lastly (the cyan line in Fig.~\ref{fig:phasediagram}), the phase transition
between the Haldane and N{\'e}el phases is the 2D Ising-type transition.  In
this case, the critical value $\Delta_{\rm l}^{({\rm c;H,N})}$ of
$\Delta_{\rm l}$ at the thermodynamic limit for a given value of
$J_{{\rm l},b}$ can be evaluated by the phenomenological renormalization-group
(PRG) method~\cite{PRGmethod} in the following way.  We calculate the
finite-size critical value $\Delta_{\rm l}^{({\rm c;H,N})}(L,L+2)$ by
solving the PRG equation given by
\begin{equation}
    L\,\Delta E_{1}^{\rm P}(L,0) = (L+2)\,\Delta E_{1}^{\rm P}(L+2,0)\,.
    \label{eq:H-N}
\end{equation}
Then, we extrapolate the finite-size results to the \hbox{$L\!\to\!\infty$}
limit to evaluate $\Delta_{\rm l}^{({\rm c;H,N})}$.  In a practical
extrapolation we fitted them to a quadratic function of $(L\!+\!1)^{-2}$ by
using the least-squares method.  In Fig.~\ref{fig:H-N} we illustrate this
procedure for the case of \hbox{$J_{{\rm l},b}\!=\!2.0$}.

Before closing this section, we make two more comments on the phase diagram
shown in Fig.~\ref{fig:phasediagram}.
{First, there are two tetracritical points, at one of which the
ferromagnetic-nematic TLL and N{\'e}el-nematic TLL phase boundary lines
come in contact with each other, and at the other the $XY$1-Haldane,
Haldane-N{\'e}el, N{\'e}el-nematic TLL, and nematic TLL-$XY$1 phase boundary
lines merge.
As regards the latter tetracritical point, we have concluded that this is the
case based on the physical consideration by referring to the result of the
perturbational calculation discussed in Sect.~3 [see
Fig.~\ref{fig:mapping}(b)].  We note that
it is hard to draw this conclusion only from the present ED calculation for 
finite-$L$ systems with up to \hbox{$L\!=\!14$} rungs.}
Secondly,
{the results of our ED calculation suggest that,}
when the value of $J_{{\rm l},b}$ is sufficiently
small (at least when \hbox{$J_{{\rm l},b}\!\lsim\!-4.0$}), the $XY$1 phase may
appear below the N{\'e}el phase.  The phase transition between these two
phases is the BKT transition with the STSB, and the phase boundary line can be
estimated from the
equation~\cite{okamoto-nomura,nomura-okamoto-1,nomura-okamoto-2}
\begin{equation}
    \Delta E_{1}^{\rm P}(L,0) = \Delta E_{0}^{\rm P}(L,1)
    \label{eq:N-XY1}
\end{equation}
instead of Eq.~(\ref{eq:N-nTLL}).  We have tried to estimate this phase
boundary line in a similar way.
{Although the finite-size critical points which satisfy
Eq.~(\ref{eq:N-XY1}) exist, the numerical error coming from the
extrapolation \hbox{$L\!\to\!\infty$} is fairly large in this case.
Thus, we believe that the $XY$1 phase exists in this region, but it is
difficult to draw the phase boundary line.  In order to overcome this
difficulty, it is indispensable to attempt more sophisticated analyses,
which are beyond the scope of the present study.}

\section{Perturbational Calculations}

In order to elucidate the characteristic features of the phase diagram shown in
Fig.~\ref{fig:phasediagram}, we have performed the first-order perturbational
calculation from the strong-rung coupling limit.  We begin with the two-spin
problem of the $j$th rung;
\begin{equation}
    \cH_\rr
    = J_\rr \left\{\Gamma_\rr \left(S_{j,a}^x S_{j,b}^x 
        + S_{j,a}^y S_{j,b}^y\right) + S_{j,a}^z S_{j,b}^z \right\}\,.
\end{equation}
The eigenstates and the eigenenergies of $\cH_\rr$ are as follows:
\begin{align}
   &\ket{\psi_1} = \,\ket{\up_{j,a}\up_{j,b}}\,, &&E_1 = {J_\rr  \over 4}\,, \\
   &\ket{\psi_2} = {\ket{\up_{j,a}\dn_{j,b} + \dn_{j,a}\up_{j,b}} \over
                                                               \sqrt{2}}\,,
     &&E_2 = {J_\rr (2\Gamma_\rr - 1) \over 4}\,, \\
   &\ket{\psi_3} = \,\ket{\dn_{j,a}\dn_{j,b}}\,, &&E_3 = {J_\rr \over 4}\,, \\
   &\ket{\psi_4} = {\ket{\up_{j,a}\dn_{j,b} - \dn_{j,a}\up_{j,b}}
                                                          \over \sqrt{2}}\,,
     &&E_4 = -{J_\rr (2\Gamma_\rr + 1) \over 4}\,,
\end{align}
where $\ket{\up_{j,\ell}}$ and $\ket{\dn_{j,\ell}}$ denote, respectively, the
\hbox{$S_{j,\ell}^z\!=\!\frac{1}{2}$} and
\hbox{$S_{j,\ell}^z\!=\!-\frac{1}{2}$} states.  
We take first the three states
$\ket{\psi_1}$, $\ket{\psi_2}$, and $\ket{\psi_3}$, and neglect the
$\ket{\psi_4}$ state, because we suppose that \hbox{$J_\rr(=\!-1)$} is
ferromagnetic.
In this restricted space, we interpret these three states as being the
\hbox{$T_j^z\!=\!+1$}, \hbox{$T_j^z\!=\!0$}, and \hbox{$T_j^z\!=\!-1$} states,
respectively, of the pseudospin operator $\Vec{T_j}$ with \hbox{$T_j\!=\!1$}.
Comparing the matrix elements, we see that the original spin operators
$S_{j,\ell}^\mu$ (\hbox{$\mu\!=\!x$},$y$,$z$) can be expressed by
\begin{equation}
    S_{j,\ell}^\mu = {1 \over 2}T_j^\mu
\end{equation}
within the restricted space.  Thus, in the first-order perturbation theory,
we can obtain the effective Hamiltonian
\begin{eqnarray}
   &&\!\!\!\!\!\!\!\!\!\!\!\!\!\!\!\cHeff
    = \Jeff 
      \left\{ \sum_{j=1}^L \left( T_j^x T_{j+1}^x + T_j^y T_{j+1}^y 
         + \Deltaeff T_j^z T_{j+1}^z \right) \right. \nonumber \\
      &&~~~~~~~~~~~~~~~\left.      + \Deff \sum_{j=1}^L (T_j^z)^2 
      \right\}\,,
    \label{eq:Heff}
    \\
    &&\Jeff = {J_{\rl,a} + J_{\rl,b} \over 4}\,,              \\
    &&\Deltaeff = \Delta_\rl\,,                               \\
    &&\Deff = {J_\rr (1-\Gamma_\rr) \over 2\Jeff}
    \,,
\end{eqnarray}
where the $\Deff$ term comes from the energy differences
\hbox{$E_1\!-\!E_2$} and \hbox{$E_3\!-\!E_2$}.  The effective Hamiltonian
(\ref{eq:Heff}) is a fundamental model for the anisotropic \hbox{$T\!=\!1$}
chain, which has been investigated by several
authors.~\cite{schulz,nijs-rommelse,chen-etal}  In particular, Chen {\it et}
{\it al.}~\cite{chen-etal} have numerically determined the phase diagram of
$\cHeff$ on the $\Deltaeff$ versus $\Deff$ plane.

\begin{figure}[t]
   \begin{center}
       \scalebox{0.25}{\includegraphics{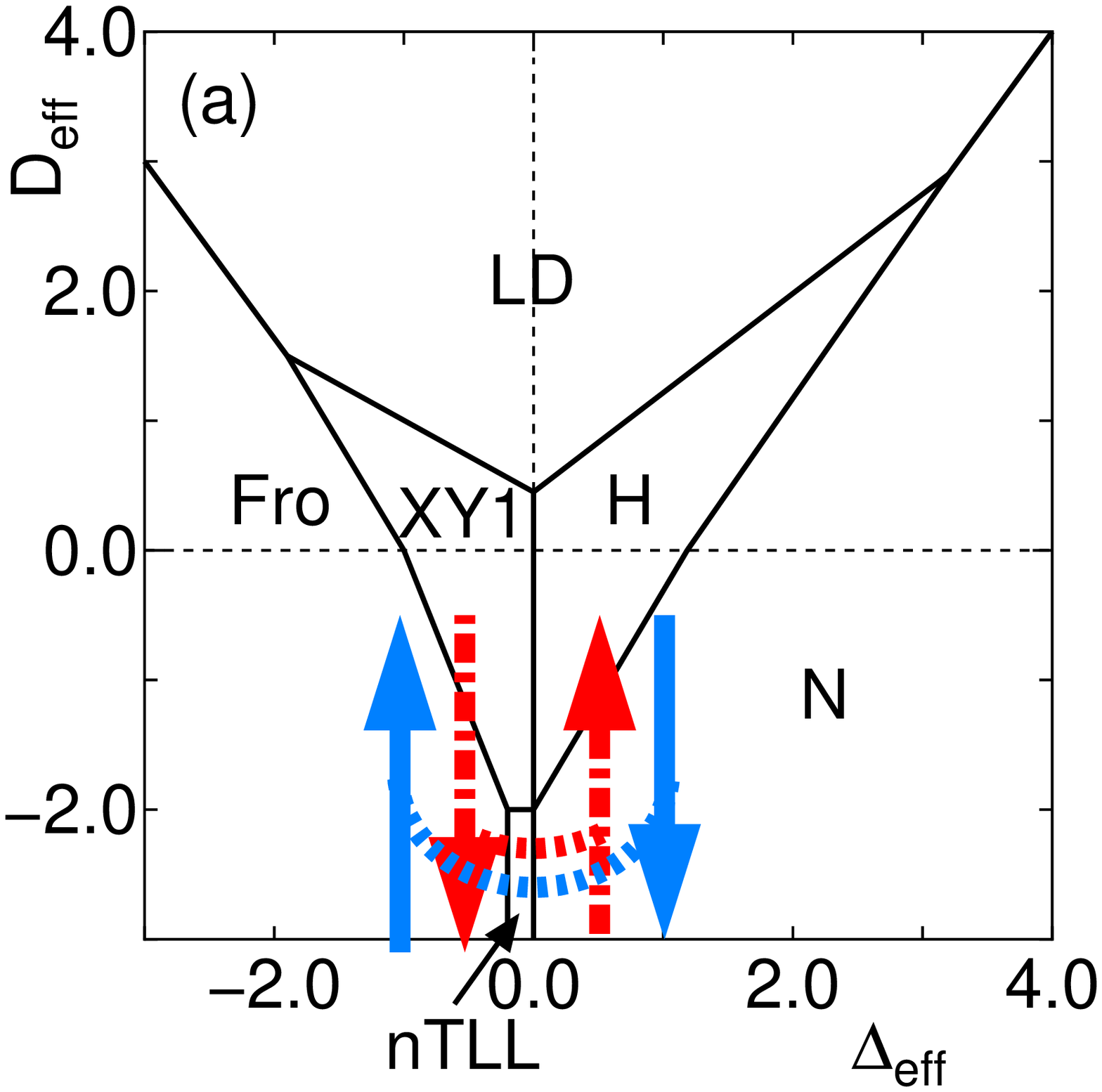}}~~
       \scalebox{0.25}{\includegraphics{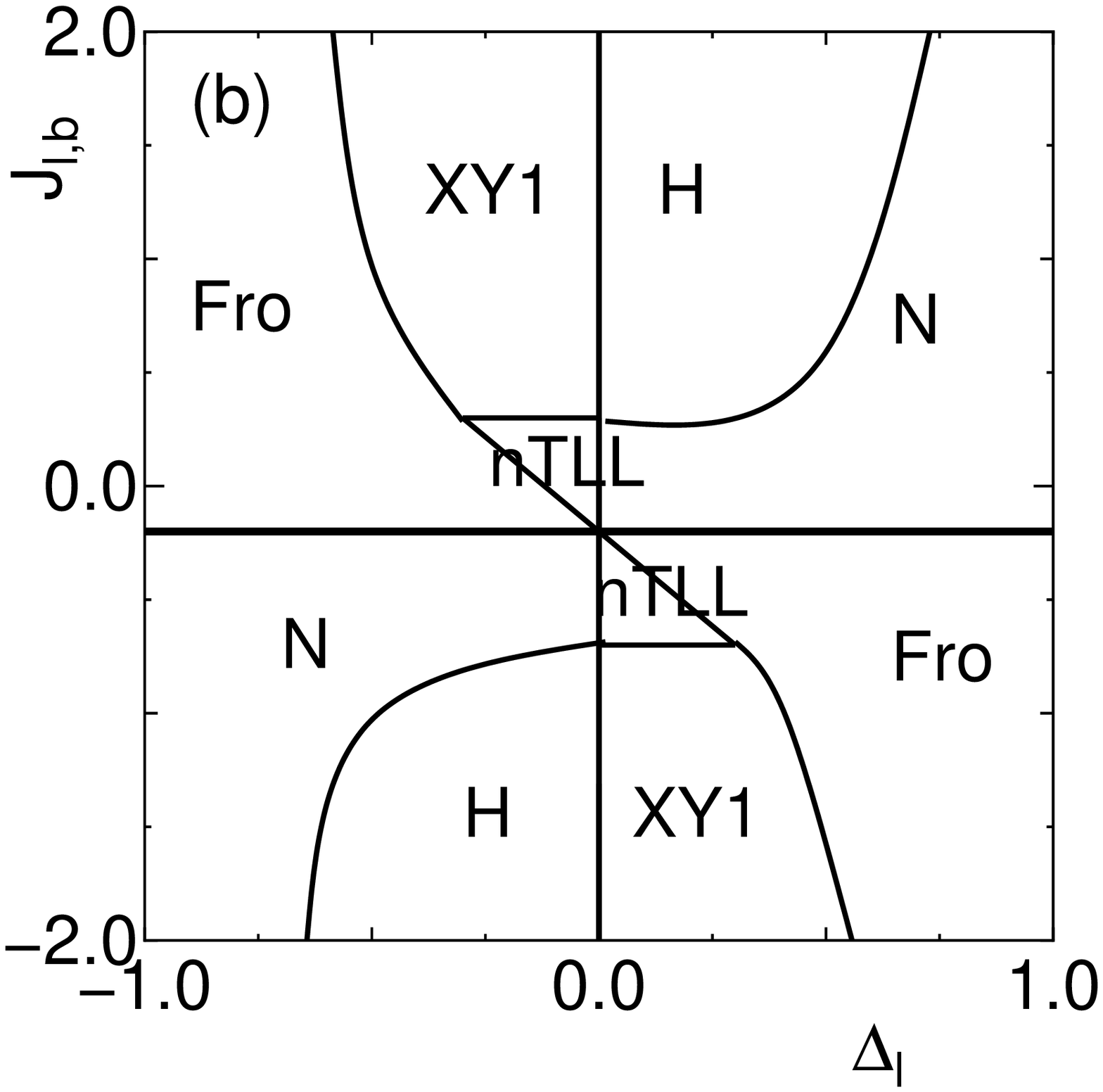}}
   \end{center}
   \caption{(Color online)
    (a) Schematic phase diagram of $\cHeff$ numerically obtained by
    Chen {\it et} {\it al.}~\cite{chen-etal}.  The regions designated by LD,
    Fro, XY1, H, N, and nTLL are those of the large-$D$, ferromagnetic, $XY1$,
    Haldane, N{\'e}el, and nematic TLL phases, respectively.
    (b) Schematic phase diagram of the present system (${\cal H}$) predicted
    by the perturbation theory based on $\cHeff$ in the same case
    (\hbox{$J_{{\rm l},a}\!=\!0.2$}, \hbox{$J_\rr\!=\!-1.0$}, and
    \hbox{$\Gamma_\rr\!=\!0.5$}) as that in Fig.~\ref{fig:phasediagram}.
    }
   \label{fig:mapping}
\end{figure}

Figure \ref{fig:mapping}(a) schematically shows the above-mentioned phase
diagram of the effective Hamiltonian $\cHeff$ obtained by Chen {\it et}
{\it al.}~\cite{chen-etal}  We have chosen that \hbox{$J_{{\rm l},a}\!=\!0.2$},
\hbox{$J_\rr\!=\!-1.0$}, and \hbox{$\Gamma_\rr\!=\!0.5$}, and therefore
the parameters in $\cHeff$ are given by
\begin{eqnarray}
    &&\Jeff = {J_{\rl,b} + 0.2 \over 2}\,,  \\  
    &&\Deltaeff = \Delta_\rl\,, \\
    &&\Deff = -{1 \over J_{\rl,b} + 0.2}\,.
\end{eqnarray}
From Fig.~\ref{fig:mapping}(a), $\cHeff$, and the above three equations,
we obtain the phase diagram given in Fig.~\ref{fig:mapping}(b).  This phase
diagram has a point symmetry with respect to the point
\hbox{$(\Delta_\rl,J_{\rl,b})\!=\!(0.0,-0.2)$}, which can be explained as
follows.  If we make the unitary transformations,
\begin{eqnarray}
    &&\!\!\!\!\!\!\!\!\!\!T_{2j}^x\to -T_{2j}^x\,,~~T_{2j}^y\to -T_{2j}^y\,,
      ~~T_{2j}^z\to T_{2j}^z\,, \\
    &&\!\!\!\!\!\!\!\!\!\!T_{2j+1}^x \to T_{2j+1}^x\,,
      ~~T_{2j+1}^y\to T_{2j+1}^y\,,~~T_{2j+1}^z\to T_{2j+1}^z\,, \nonumber \\
\end{eqnarray}
the effective Hamiltonian $\cHeff$ is transformed into
\begin{eqnarray}
    &&\!\!\!\!\!\!\!\!\!\!\!\!\cHeff
    \to -\Jeff 
      \left\{ \sum_{j=1}^L \left( T_j^x T_{j+1}^x + T_j^y T_{j+1}^y
                - \Deltaeff T_j^z T_{j+1}^z \right) \right. \nonumber \\
      &&~~~~~~~~~~~~~~~~~~~~~\left.      - \Deff \sum_{j=1}^L (T_j^z)^2 
      \right\}\,.
\end{eqnarray}
Thus, we see that the two points, $(\Jeff,\Deltaeff,\Deff)$ and
$(-\Jeff,-\Deltaeff,-\Deff)$, belong to the same phase.  The transformation
\begin{equation}
    (\Jeff,\Deltaeff,\Deff) \to (-\Jeff,-\Deltaeff,-\Deff)
\end{equation}
can be carried out by
\begin{eqnarray}
   &&J_{\rl,b} + 0.2 \to -(J_{\rl,b} + 0.2)\,, \\
   &&\Delta_\rl \to -\Delta_\rl\,,
\end{eqnarray}
which explains the point symmetry of the phase diagram shown in
Fig.~\ref{fig:mapping}(b).

The above arguments concerning the point symmetry well explain the reason
why the ferromagnetic, N{\'e}el, and nematic TLL phases appear in two places
in Fig.~\ref{fig:phasediagram}(a).  They suggest that the $XY1$ state appears
as the ground state on the lower outside of Fig.~\ref{fig:phasediagram}(a)
(\hbox{$J_{{\rm l},b}\!<\!-2.0$}).
Although we believe the existence of the $XY1$ phase,
we are unfortunately not able to determine its phase boundary,
as already noted in the last paragraph in Sect.~2.
Also, the Haldane phase does not appear in the region of negative
$J_{{\rm l},b}$ in Fig.~\ref{fig:phasediagram}(a), and further the partial
ferrimagnetic phase found in Fig.~\ref{fig:phasediagram}(a) does not appear
in Fig.~\ref{fig:mapping}(a).  We think that the appearance of the partial
ferrimagnetic phase is attributed to the strong frustration effect, which is
not taken into account in the present first-order perturbational calculation.

Let us explain the correspondence between Fig.~\ref{fig:phasediagram}(a) and
Fig.~\ref{fig:mapping}(a) a little bit more in detail.  When $J_{\rl,b}$
changes, on the vertical line \hbox{$\Delta_\rl\!=\!1.0$} in
Fig.~\ref{fig:phasediagram}(a), from a large positive value to $-0.2$, the
parameter $\Jeff$ is always positive, \hbox{$\Deltaeff\!=\!1.0$}, and
$\Deff$ changes from a small negative value to a large negative value.  Thus,
the corresponding point in Fig.~\ref{fig:mapping}(a) moves along the
downward-arrowed {blue solid line}.  On the other hand, when
$J_{\rl,b}$
changes, again on the vertical line \hbox{$\Delta_\rl\!=\!1.0$} in
Fig.~\ref{fig:phasediagram}(a), from $-0.2$ to a large negative value, the
parameter $\Jeff$ is always negative, \hbox{$\Deltaeff\!=\!1.0$}, and
$\Deff$ changes from a large positive value to a small positive value.  Since
the points $(\Jeff,\Deltaeff,\Deff)$ and $(-\Jeff,-\Deltaeff,-\Deff)$ belong to
the same phase as already explained, these changes in the parameters in
$\cHeff$ are equivalent to those in the case where $\Jeff$ is positive,
\hbox{$\Deltaeff\!=\!-1.0$}, and $\Deff$ changes from a large negative value to
a small negative value.  Namely, the corresponding point in
Fig.~\ref{fig:mapping}(a) moves along the upward-arrowed
{blue solid line}.
Although the two {blue solid lines}
are separated from each other, we surmise
that they are connected with each other by the
{blue dotted line} in the higher-order perturbation theory
at least near \hbox{$\Deltaeff\!=\!0.0$}.
{Another possibility is that they are connected through
\hbox{$\Deltaeff\!=\!\pm\infty$.}
Thus, the phase of the original system ${\cal H}$
successively changes to the N{\'e}el, nematic TLL, and ferromagnetic phases
with decreasing $J_{\rl,b}$, which well explains the phases on the
\hbox{$\Delta_\rl\!=\!1.0$} line in Fig.~\ref{fig:phasediagram}(a).  We can
make a very similar discussion, for example, for the line
\hbox{$\Delta_\rl\!=\!-0.5$} in Fig.~\ref{fig:phasediagram}(a) by use of the
{red dot-dashed lines} 
in Fig.~\ref{fig:mapping}(a).  Namely, the phase successively
changes to the $XY$1, ferromagnetic, nematic TLL, and N{\'e}el phases with
decreasing $J_{\rl,b}$ in this case.  Thus, the perturbational calculation
together with the numerical result obtained by
Chen {\it et} {\it al.}~\cite{chen-etal} qualitatively explains the
numerically determined phase diagram of the present system (${\cal H}$).

{Here, we discuss the validity of the perturbational calculation.  At a glance,
this perturbational calculation is thought to be valid for the strong-rung
coupling case, \hbox{$|J_{\rl,b}|\!\ll\!|J_\rr|$}.  However, the bosonization
theory from the weak-rung coupling limit shows that the rung coupling is
relevant and is renormalized to the strong-rung coupling case, as is known in
the simple ladder case.  This fact strongly suggests that the present
perturbational calculation is qualitatively valid even for the weak-rung
coupling case.}

\section{DMRG Calculations}

In this section, we present several results of the DMRG calculation, which
supplement the reliability of the phase diagram shown in
Fig.~\ref{fig:phasediagram}.  In this calculation, we assume the open
boundary condition (OBC) for the Hamiltonian ${\cal H}$, where the sum over
$j$ is taken from \hbox{$j\!=\!1$} to \hbox{$j\!=\!L\!-\!1$}.  We treat the
finite-size systems with up to \hbox{$L\!=\!96$} rungs, $L$ being assumed to
be even. The number of block states in the DMRG calculation required to achieve
the desired accuracy depends on the state targeted.  In our calculation, we
have kept up to 600 block states for the most severe case and checked that the
results are accurate enough for our arguments below by monitoring
the dependences of the data and the truncation error on the number of kept
states.

The physical quantities
which we calculate and analyze are the energy gaps $\Delta E^{\rm O}(L,1)$ and
$\Delta E^{\rm O}(L,2)$, the local magnetization $m_{\ell}^{\rm O}(j;L,M)$,
the two-spin correlation function $\omega_{\ell}^{\mu\mu,{\rm O}}(j;L,M)$
(\hbox{$\mu\!=\!x,z$}), and the nematic four-spin correlation function
$\omega_{\ell}^{++--,{\rm O}}(j;L,M)$.  Among these quantities,
$\Delta E^{\rm O}(L,1)$ and $\Delta E^{\rm O}(L,2)$ are defined, respectively,
by
\begin{eqnarray}
  && \!\!\!\!\!\!\!\!\!\!
  \Delta E^{\rm O}(L,1)=E_0^{\rm O}(L,1)-E_0^{\rm O}(L,0)\,, \\
  && \!\!\!\!\!\!\!\!\!\!
  \Delta E^{\rm O}(L,2)=E_0^{\rm O}(L,2)-E_0^{\rm O}(L,0)\,,
\end{eqnarray}
where $E_0^{\rm O}(L,M)$ is the lowest-energy eigenvalue of
${\cal H}$ under the OBC within the subspace determined by $L$ and $M$.  It
is noted that except for the ferromagnetic and partial ferrimagnetic phases,
$E_0^{\rm O}(L,0)$ gives the ground-state energy.  Further,
$m_{\ell}^{\rm O}(j;L,M)$ is defined by
\begin{equation}
   m_{\ell}^{\rm O}(j;L,M) = \bigl\langle S_{j,\ell}^z \bigr\rangle_{L,M}\,,
\end{equation}
and also $\omega_{\ell}^{\mu\mu,{\rm O}}(j;L,M)$ and
   $\omega_{\ell}^{++--,{\rm O}}(j;L,M)$ are defined, respectively, by
\begin{equation}
  \omega_{\ell}^{\mu\mu,{\rm O}}(j;L,M)
     =\bigl\langle S_{j_0,\ell}^\mu\,S_{j_0+j,\ell}^\mu 
        \bigr\rangle_{L,M} \,,
\end{equation}
\begin{equation}
  \!\!\!
  \omega^{++--,{\rm O}}(j;L,M)
     =\bigl\langle S_{j_0,a}^{+}\,S_{j_0,b}^{+}\,
         S_{j_0+j,a}^{-}\,S_{j_0+j,b}^{-} \bigr\rangle_{L,M}\,,
\end{equation}
where depending upon whether \hbox{$j(>\!0)$} is odd or even,
\hbox{$j_0\!=\!\frac{L+1-j}{2}$} or \hbox{$j_0\!=\!\frac{L-j}{2}$},
respectively,
which means that the two rungs $j_0$ and $j_0\!+\!j$ are at (almost) equal
distances from the center of the open ladder.  In the above three equations,
$\langle\cdots \rangle_{L,M}$ denotes the expectation value associated with
the lowest-energy state within the subspace determined by $L$ and $M$ in the
system under the OBC; when \hbox{$M\!=\!M_{\rm g}(L)$}, this expectation value
is simply the ground-state expectation value for the system with $L$ rungs.

\begin{figure}[t]
   \begin{center}
       \scalebox{0.45}{\includegraphics{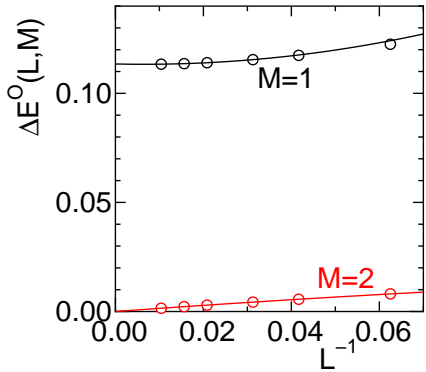}}
   \end{center}
   \caption{(Color online)
   Dependences of $\Delta E^{\rm O}(L,1)$ and
   $\Delta E^{\rm O}(L,2)$ on $L^{-1}$ at the point
   $(\Delta_{\rm l}\!=\!0.5,J_{{\rm l},b}\!=\!-0.5)$ (in the nematic TLL phase)
   in the phase diagram
   shown in Fig.~\ref{fig:phasediagram}.  The solid lines represent the
   least-squares fittings, where the data for \hbox{$L\!=\!32$}, $48$, $64$,
   $96$ are used.
   }
   \label{fig:gapNTLL}
\end{figure}

\begin{figure}[t]
   \begin{center}
       \scalebox{0.22}{\includegraphics{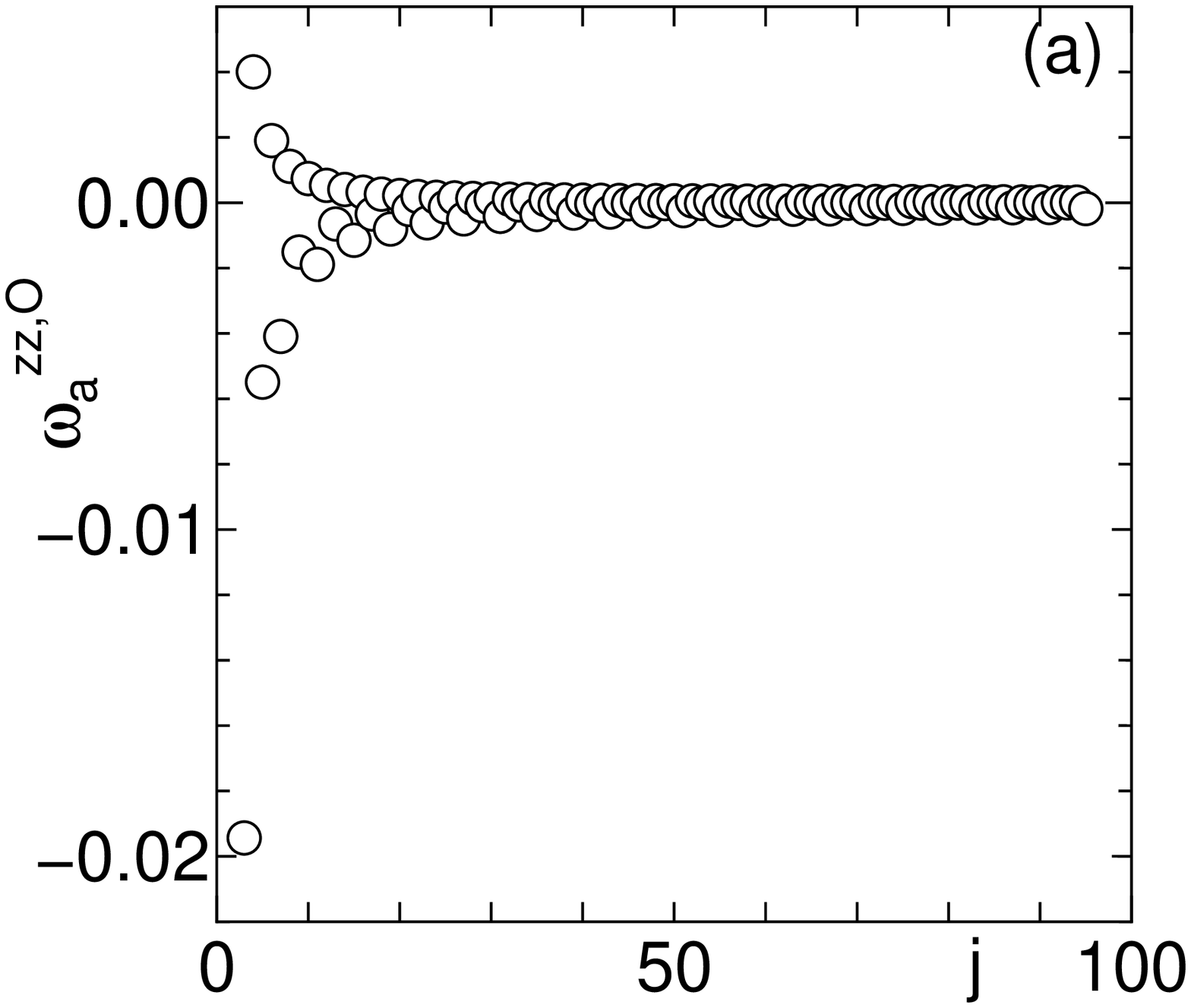}}~~
       \scalebox{0.22}{\includegraphics{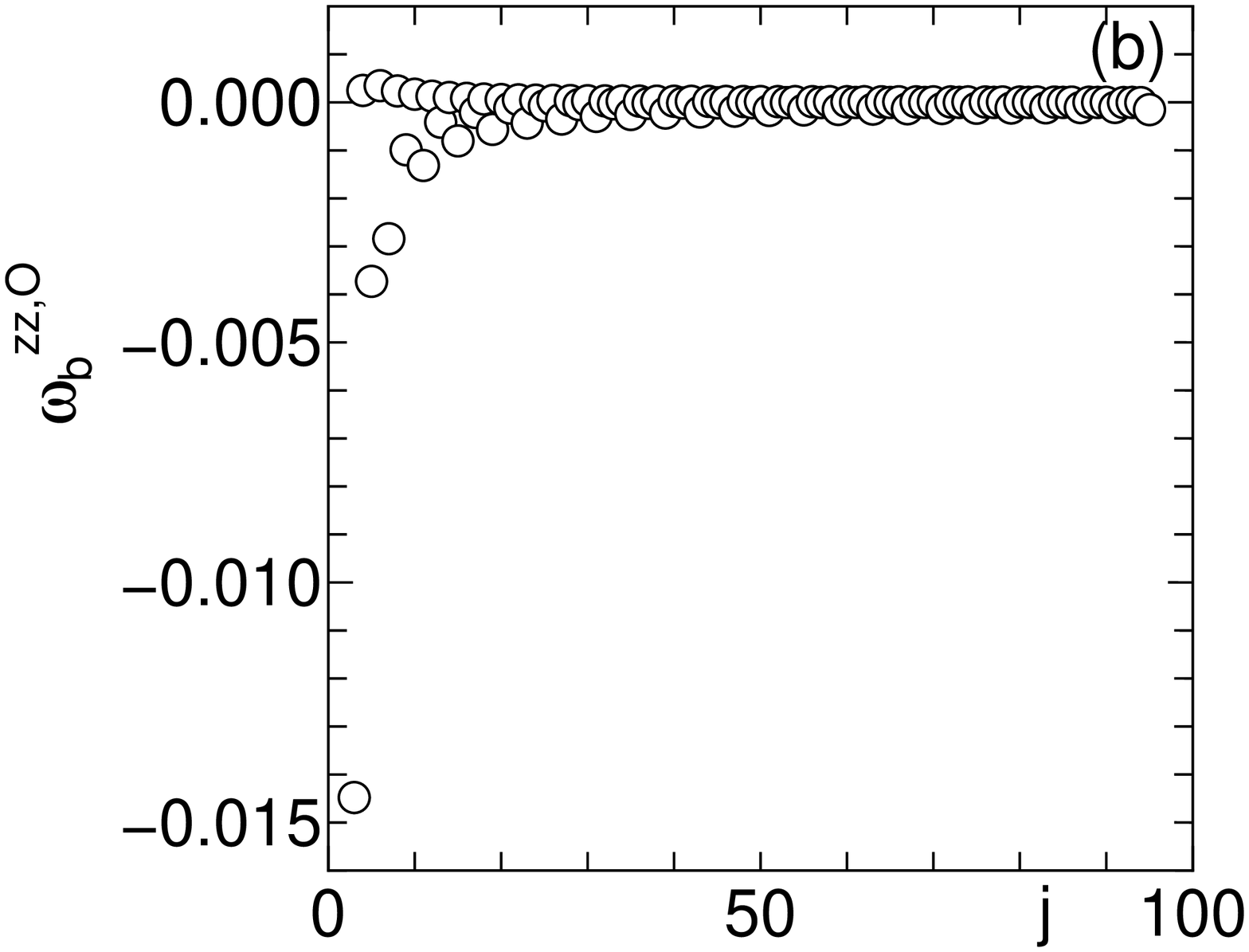}}
       \scalebox{0.22}{\includegraphics{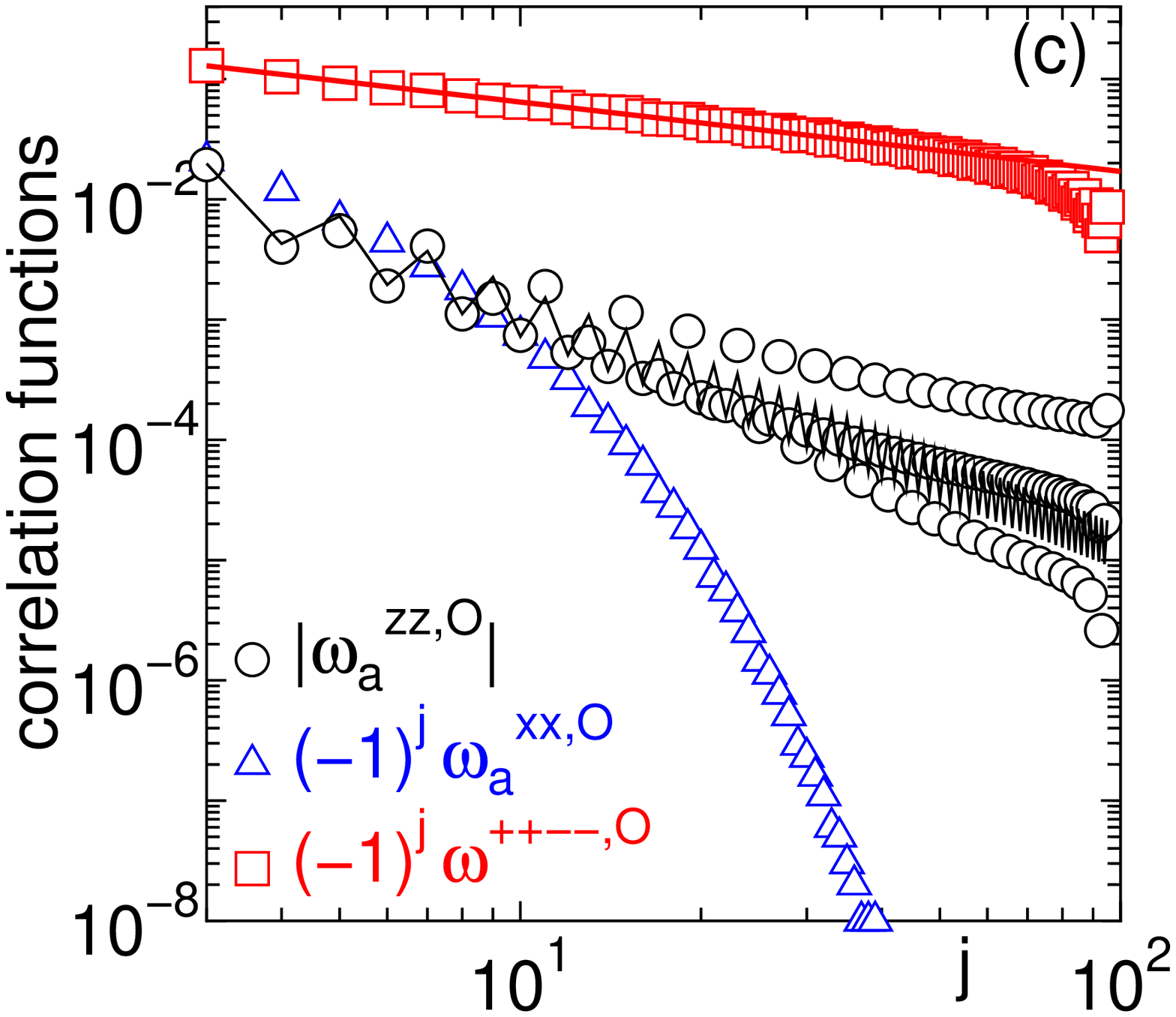}}~~
       \scalebox{0.22}{\includegraphics{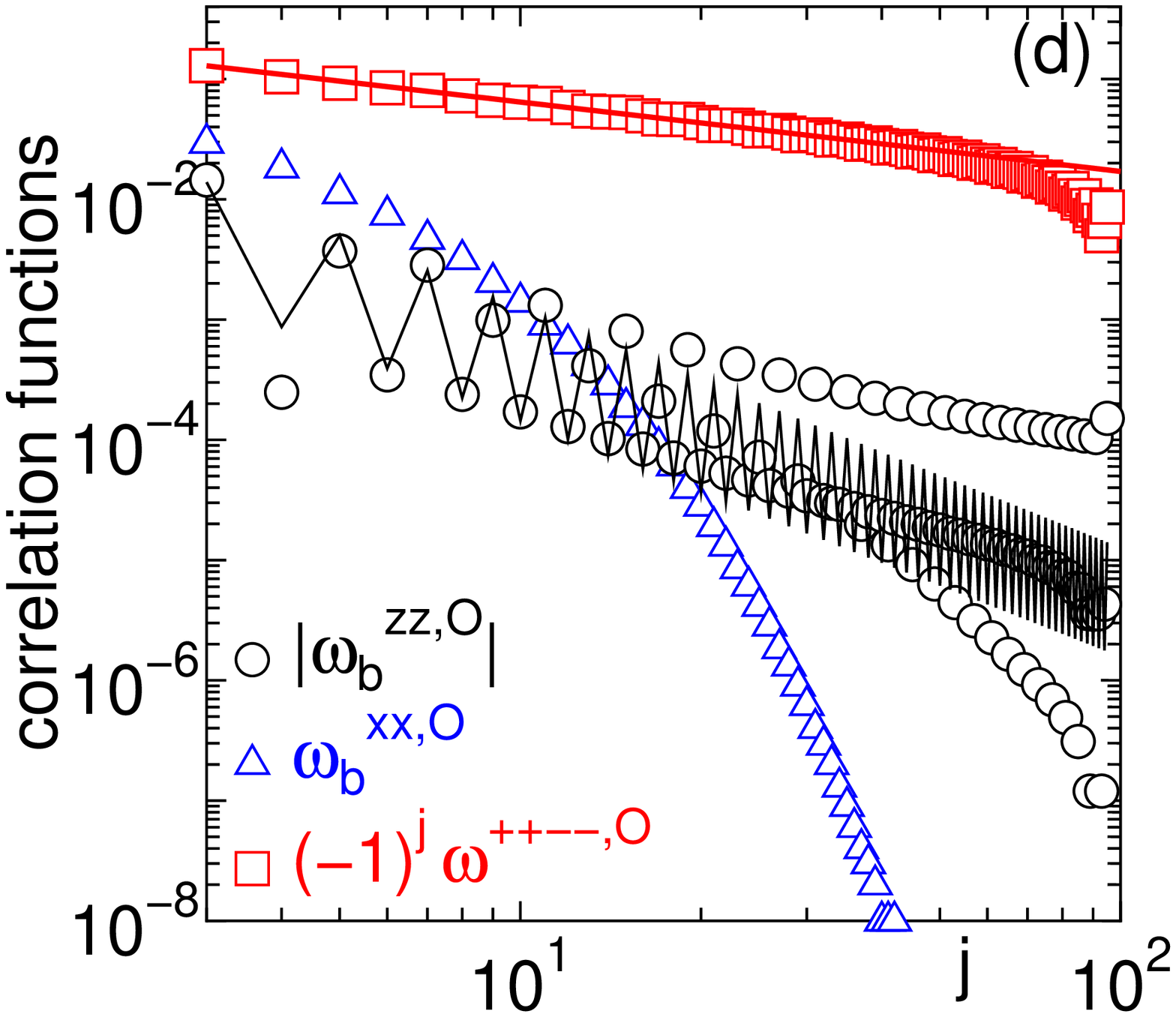}}
   \end{center}
   \caption{(Color online)
Correlation functions at the point
$($\hbox{$\Delta_{\rm l}\!=\!0.5$}$,J_{{\rm l},b}\!=\!-0.5)$ (in the nematic
TLL phase).  In (a) and (b), the $j$-dependences of
$\omega_{\ell}^{zz,{\rm O}}(j;96,0)$ for \hbox{$\ell\!=\!a$} and
\hbox{$\ell\!=\!b$}, respectively, are shown in a linear scale.  In (c),
$|\omega_{a}^{zz,{\rm O}}(j;96,0)|$, $(-1)^j \omega_{a}^{xx,{\rm O}}(j;96,0)$,
and $(-1)^j \omega^{++--,{\rm O}}(j;96,0)$ are shown in a log-log scale, while
in (d), $|\omega_{b}^{zz,{\rm O}}(j;96,0)|$, $\omega_{b}^{xx,{\rm O}}(j;96,0)$,
and $(-1)^j \omega^{++--,{\rm O}}(j;96,0)$ are shown in a log-log scale.
[Note that the same data of $(-1)^j \omega^{++--,{\rm O}}(j;96,0)$ are shown in
both (c) and (d) for comparison.]  The solid lines in (c) and (d) represent the
least-squares fitting of
$\omega_{\ell}^{zz,{\rm O}}(j;96,0)$ and $\omega^{++--,{\rm O}}(j;96,0)$ to
Eqs.~(\ref{eq:C_spin_zz}) and (\ref{eq:C_nem}), respectively, where the data
from \hbox{$j\!=\!6$} to \hbox{$j\!=\!36$} are used.
}
   \label{fig:corfnNTLL}
\end{figure}

Figure~\ref{fig:gapNTLL} shows our DMRG results for the $1/L$-dependence of
the energy gaps $\Delta E^{\rm O}(L,1)$ and $\Delta E^{\rm O}(L,2)$ at the
point $(\Delta_{\rm l}\!=\!0.5,J_{{\rm l},b}\!=\!-0.5)$, which belongs to the
nematic-TLL-phase region in the phase diagram shown in
Fig.~\ref{fig:phasediagram}.  Performing the \hbox{$L\!\to\!\infty$}
extrapolation of these gaps, $\Delta E^{\rm O}(L,M)$'s, we have fitted them to
quadratic functions of $L^{-1}$ by use of the least-squares method.  The solid
lines in Fig.~\ref{fig:gapNTLL}, which show these extrapolations, demonstrate
that at the thermodynamic (\hbox{$L\!\to\!\infty$}) limit, the former gap
is finite [\hbox{$\Delta E^{\rm O}(1)\!\sim\!0.113$}], while the latter gap
vanishes [\hbox{$\Delta E^{\rm O}(2)\!\sim\!6\!\times\!10^{-6}$}].  The
binding energy of two magnons is given by
\hbox{$E_{\rm betm}\!=\!2\Delta E^{\rm O}(1)\!-\!\Delta E^{\rm O}(2)
\!\sim\!0.227$} and is definitely finite, which is consistent with the fact
that the nematic TLL state accompanies the two-magnon bound state.  In
Fig.~\ref{fig:corfnNTLL} we present the DMRG results for the correlation
functions for \hbox{$L\!=\!96$} and \hbox{$M\!=\!0$} at the point
(\hbox{$\Delta_{\rm l}\!=\!0.5$}, \hbox{$J_{{\rm l},b}\!=\!-0.5$}).  We note
that the four-rung-period oscillation as a function of $j$ in the correlation
function $\omega_{\ell}^{zz,{\rm O}}(j;96,0)$ comes from the open boundary
effect.\cite{four-rung-period}  From the figure, we see that the transverse
two-spin correlation function $\omega_{\ell}^{xx,{\rm O}}(j;96,0)$ decays
exponentially, while the longitudinal two-spin correlation function
$\omega_{\ell}^{zz,{\rm O}}(j;96,0)$ and the nematic four-spin correlation
function $\omega^{++--,{\rm O}}(j;96,0)$ decay algebraically.  Here, as is
expected, $\omega^{++--,{\rm O}}(j;96,0)$ is more dominant than
$\omega_{\ell}^{zz,{\rm O}}(j;96,0)$.  The asymptotic forms of the latter two
correlation functions in the nematic TLL phase are expected to
be\cite{bosonization-text}
\begin{eqnarray}
  && \!\!\!\!\!\!\!\!\!\!
  \omega_{\ell}^{zz,{\rm O}}(j;L \to \infty,0)
      \sim (-1)^j \frac{\alpha_{1,\ell}}{j^{\,\eta_\ell^{zz}}}
                                + \frac{\alpha_{2,\ell}}{j^{\,2}}         \,,
\label{eq:C_spin_zz}\\
  && \!\!\!\!\!\!\!\!\!\!
  \omega^{++--,{\rm O}}(j;L \to \infty,0)
      \sim (-1)^j \frac{\alpha}{j^{\,\eta^{++--}}}  \,,
\label{eq:C_nem}
\end{eqnarray}
where $\alpha_{1,\ell}$, $\alpha_{2,\ell}$, and $\alpha$ are numerical
constants.  The decay exponents obey the relation
$\eta_\ell^{zz} \eta^{++--}=1$.  We have performed the least-squares fitting of
the nematic four-spin correlation function $\omega^{++--,{\rm O}}(j;96,0)$ to
Eq.~(\ref{eq:C_nem}) using the data for \hbox{$6\!\le\!j\!\le\!36$}.  We have
thereby obtained the estimate of the decay exponent,
\hbox{$\eta^{++--}\!\sim\! 0.577$}, which is much smaller than 1 and supports
the prediction of the dominant nematic correlation.  For the longitudinal
two-spin correlations, we have unfortunately not been able to achieve a
precise estimation of
the decay exponent due to rather large open-boundary effects.
We have fitted $\omega_{\ell}^{zz,{\rm O}}(j;96,0)$ to
Eq.~(\ref{eq:C_spin_zz}) using the data for several ranges of $j$, namely,
\hbox{$3\!\le\!j\!\le\!20$}, \hbox{$4\!\le\!j\!\le\!24$},
\hbox{$5\!\le\!j\!\le\!30$}, and  \hbox{$6\!\le\!j\!\le\!36$}.
Then, we have obtained the estimates in the ranges of
\hbox{$1.70\!\le\!\eta_a^{zz}\!\le\!2.11$} and
\hbox{$1.63\!\le\!\eta_b^{zz}\!\le\!2.31$}.  Although these estimates show
rather large variations, the values are consistent with the expectation
$\eta_\ell^{zz} = 1/\eta^{++--}$, namely, 
\hbox{$\eta_\ell^{zz}\!\sim\!1/0.577\!=\!1.73$}, and indicate that the
longitudinal spin correlation decays much faster than the nematic correlation.
The DMRG data thereby support the result of the ED analysis that the system at
the point $($\hbox{$\Delta_{\rm l}\!=\!0.5$}$,J_{{\rm l},b}\!=\!-0.5)$ belongs
to the nematic TLL phase.  It is also emphasized that
$\eta_a^{zz}$ and $\eta_b^{zz}$ are almost equal to each other in spite of the
fact that $J_{{\rm l},a}$ is positive, while $J_{{\rm l},b}$ is negative.
{
This result reflects the fact that the nematic TLL is a one-component
TLL,\cite{vekua-etal,hikihara-etal-1} which leads to the same decay exponents
of the correlation functions within the $a$ leg and within the $b$ leg.}

\begin{figure}[t]
   \begin{center}
       \scalebox{0.45}{\includegraphics{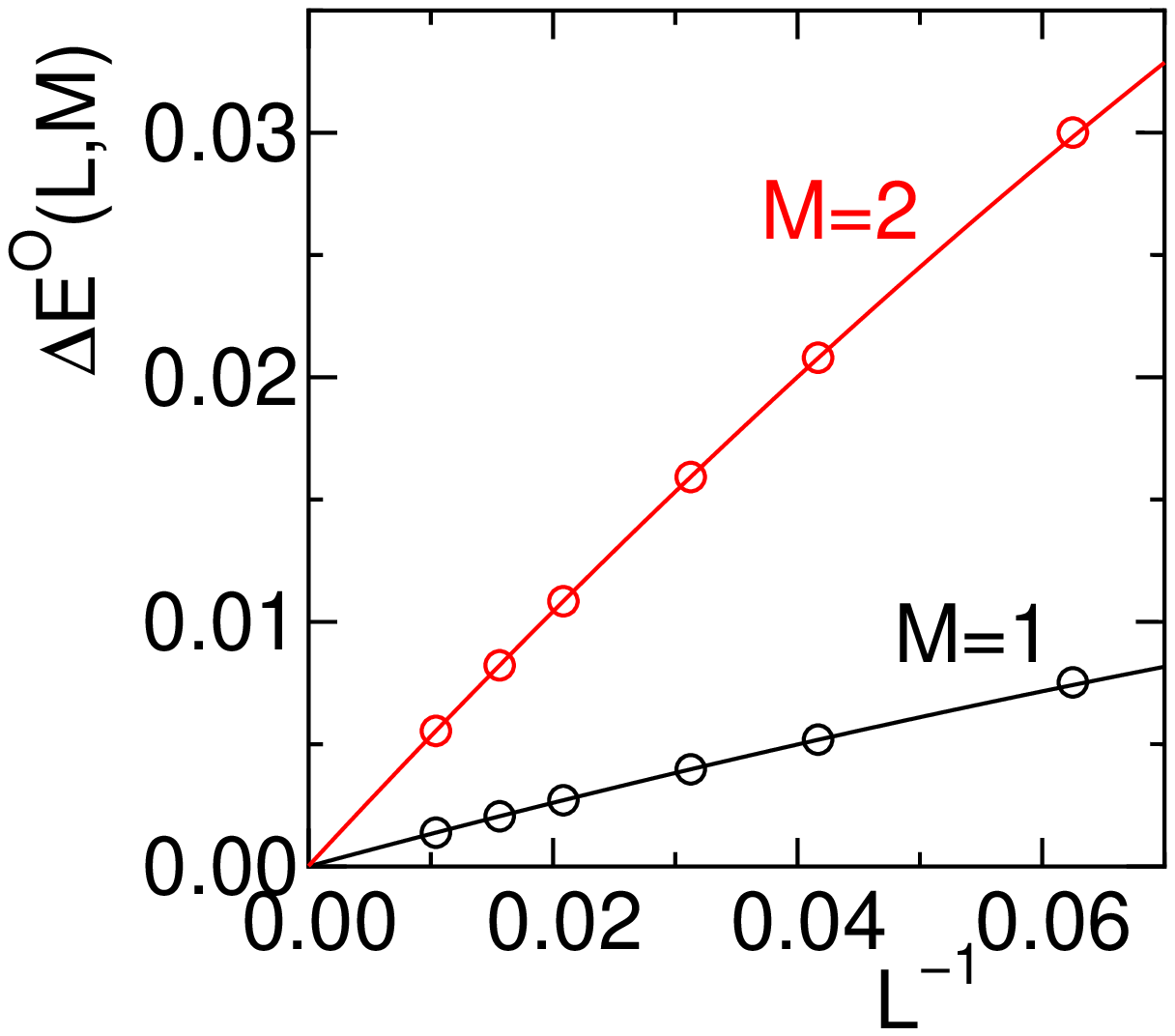}}
   \end{center}
   \caption{(Color online)
   Dependences of $\Delta E^{\rm O}(L,1)$ and
   $\Delta E^{\rm O}(L,2)$ on $L^{-1}$ at the point
   $(\Delta_{\rm l}\!=\!-0.25,J_{{\rm l},b}\!=\!1.0)$ (in the $XY$1 phase) in
   the phase diagram shown in Fig.~\ref{fig:phasediagram}.  The solid lines
   represent the least-squares fittings, where the data for \hbox{$L\!=\!32$},
   $48$, $64$, $96$ are used.
   }
   \label{fig:gapXY1}
\end{figure}

\begin{figure}[t]
   \begin{center}
       \scalebox{0.22}{\includegraphics{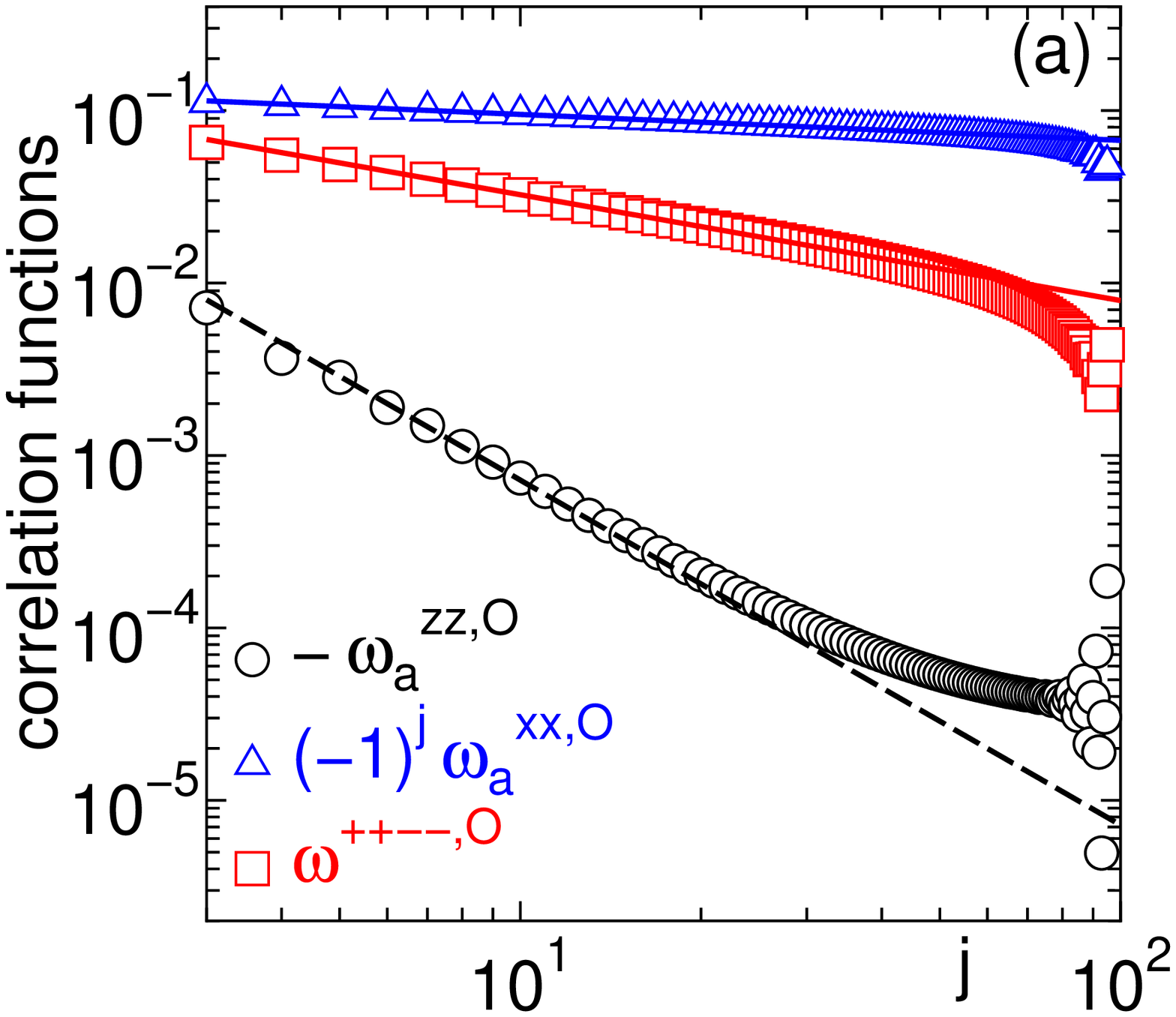}}~~
       \scalebox{0.22}{\includegraphics{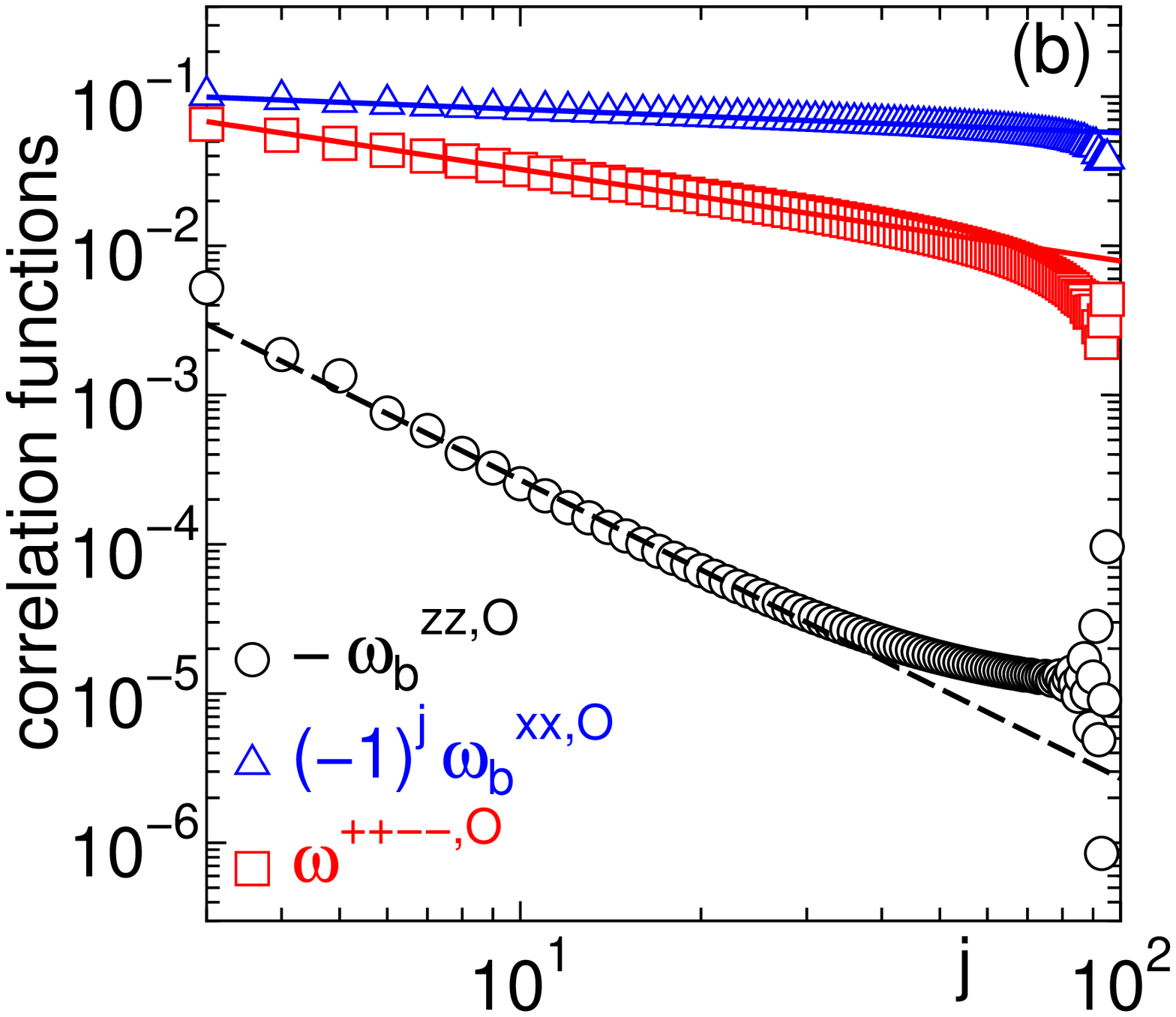}}
   \end{center}
   \caption{(Color online)
   Correlation functions at the point
   $($\hbox{$\Delta_{\rm l}\!=\!-0.25$}$,J_{{\rm l},b}\!=\!1.0)$ (in the $XY$1
   phase).  In (a) and (b), $-\omega_{\ell}^{zz,{\rm O}}(j;96,0)$ and
   $(-1)^j \omega_{\ell}^{xx,{\rm O}}(j;96,0)$ for \hbox{$\ell\!=\!a$} and
   \hbox{$\ell\!=\!b$}, respectively, are shown in a log-log scale, while
   $\omega^{++--,{\rm O}}(j;96,0)$ is shown in both (a) and (b) for comparison.
   The solid lines represent the least-squares fittings of
   $\omega_{\ell}^{xx,{\rm O}}(j;96,0)$ and $\omega^{++--,{\rm O}}(j;96,0)$ to
   Eqs.~(\ref{eq:C_spin_xx}) and (\ref{eq:C_nem_uni}), respectively, where
   the data from \hbox{$j\!=\!6$} to \hbox{$j\!=\!36$} are used.  The dashed
   lines show the power-law decay behavior, \hbox{$\sim\!1/j^2$}, of
   $\omega_{\ell}^{zz,{\rm O}}(j;96,0)$, where the first term in
   Eq.~(\ref{eq:C_spin_zz}) is neglected.
   }
   \label{fig:corfnXY1}
\end{figure}

Figure~\ref{fig:gapXY1} displays the $1/L$-dependences of the energy gaps
$\Delta E^{\rm O}(L,1)$ and $\Delta E^{\rm O}(L,2)$ at the point
$(\Delta_{\rm l}\!=\!-0.25,J_{{\rm l},b}\!=\!1.0)$, which belongs to the
$XY$1-phase region in the phase diagram shown in
Fig.~\ref{fig:phasediagram}.  We have estimated the limiting
(\hbox{$L\!\to\!\infty$}) values of these energy gaps in the same way as that
in the case of the point
(\hbox{$\Delta_{\rm l}\!=\!0.5$}, \hbox{$J_{{\rm l},b}\!=\!-0.5$})
(see the solid lines in Fig.~\ref{fig:gapXY1}), and obtained the results that
both gaps vanish [\hbox{$\Delta E^{\rm O}(1)\!\sim\!5\!\times\!10^{-6}$} and
\hbox{$\Delta E^{\rm O}(2)\!\sim\!2\!\times\!10^{-5}$}].  The DMRG results
for the correlation functions for \hbox{$L\!=\!96$} and
\hbox{$M\!=\!0$} at the point
(\hbox{$\Delta_{\rm l}\!=\!-0.25$}, \hbox{$J_{{\rm l},b}\!=\!1.0$}) are
depicted in
Fig.~\ref{fig:corfnXY1}.  In this case, all of the three correlation functions
decay algebraically.  It is expected\cite{bosonization-text} that the
asymptotic form of \hbox{$\omega_{\ell}^{zz,{\rm O}}(j;L\!\to\!\infty,0)$} in
the $XY$1 phase is again given by Eq.~(\ref{eq:C_spin_zz}), while those of
\hbox{$\omega_{\ell}^{xx,{\rm O}}(j;L\!\to\!\infty,0)$} and
\hbox{$\omega^{++--,{\rm O}}(j;L\!\to\!\infty,0)$} in this phase are written as
\begin{eqnarray}
&&   \omega_{\ell}^{xx,{\rm O}}(j;L \to \infty,0)
            \sim (-1)^j \frac{\alpha'_\ell}{j^{\,\eta_\ell^{xx}}}     \,,
\label{eq:C_spin_xx} \\
&&   \omega^{++--,{\rm O}}(j;L \to \infty,0)
         \sim \frac{\alpha'}{j^{\,\eta^{++--}}}                       \,,
\label{eq:C_nem_uni}
\end{eqnarray}
where $\alpha'_\ell$ and $\alpha'$ are constants.
It is also expected that the longitudinal and transverse spin correlations are
dual to each other, resulting in the relation 
\hbox{$\eta_\ell^{xx} \eta_\ell^{zz}\!=\!1$}, while the exponent of the nematic
correlation function is related to that of the transverse-spin correlation
function as \hbox{$\eta^{++--}\!=\!4\eta_\ell^{xx}$}.  We have estimated
$\eta_\ell^{xx}$ and $\eta^{++--}$ by the least-squares fitting of
$\omega_{\ell}^{xx,{\rm O}}(j;96,0)$ and $\omega^{++--,{\rm O}}(j;96,0)$ to
Eqs.~(\ref{eq:C_spin_xx}) and (\ref{eq:C_nem_uni}), respectively, using the
numerical data for \hbox{$6\!\le\!j\!\le\!36$}.  The results are
\hbox{$\eta_a^{xx}\!\sim\! 0.15$}, \hbox{$\eta_b^{xx}\!\sim\! 0.16$}, and
\hbox{$\eta^{++--}\!\sim\! 0.61$}, which support the expectation that
\hbox{$\eta^{++--}\!=\!4\eta_\ell^{xx}$}.  For the longitudinal-spin
correlation function, we have found that $\omega_{\ell}^{zz,{\rm O}}(j;96,0)$
is negative for all $j$ and that the oscillating component is far from sizable,
indicating that the second (uniform) term in Eq.~(\ref{eq:C_spin_zz}) is much
more dominant than the first (oscillating) term.  This observation, with our
estimate \hbox{$\eta_\ell^{xx}\!\sim\! 0.15$\,--\,$0.16$} above, is found to be
consistent with the expectation that $\eta_\ell^{xx} \eta_\ell^{zz} = 1$, since
$\eta_\ell^{zz} = 1/\eta_\ell^{xx} \sim 6.3$\,--\,$6.7$ is much larger than $2$,
the exponent of the uniform component.

\begin{figure}[t]
   \begin{center}
       \scalebox{0.45}{\includegraphics{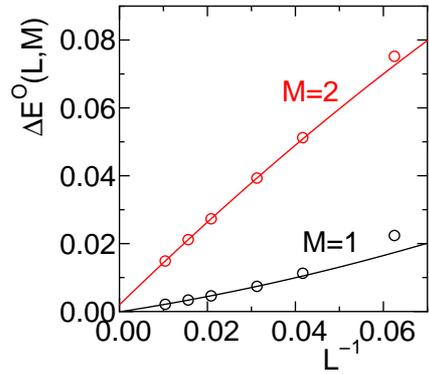}}
   \end{center}
   \caption{(Color online)
   Dependences of $\Delta E^{\rm O}(L,1)$ and
   $\Delta E^{\rm O}(L,2)$ on $L^{-1}$ at the point
   $(\Delta_{\rm l}\!=\!0.25,J_{{\rm l},b}\!=\!1.0)$ (in the Haldane phase) in
   the phase diagram shown in Fig.~\ref{fig:phasediagram}.  The solid lines
   represent the least-squares fittings where the data for \hbox{$L\!=\!32$},
   $48$, $64$, $96$ are used.
   }
   \label{fig:gapHAL}
\end{figure}

\begin{figure}[t]
   \begin{center}
       \scalebox{0.45}{\includegraphics{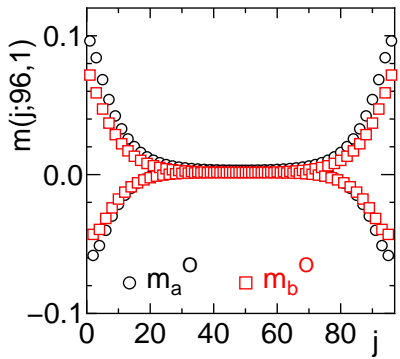}}
   \end{center}
   \caption{(Color online)
   Dependence of $m_{\ell}^{\rm O}(j;96,1)$ on $j$ at the point
   $(\Delta_{\rm l}\!=\!0.25,J_{{\rm l},b}\!=\!1.0)$ (in the Haldane phase).
   }
   \label{fig:mHAL}
\end{figure}

Figure~\ref{fig:gapHAL} shows the $1/L$-dependences of the energy gaps
$\Delta E^{\rm O}(L,1)$ and $\Delta E^{\rm O}(L,2)$ at the point
$({\Delta_{\rm l}\!=\!0.25},{J_{{\rm l},b}\!=\!1.0})$, which belongs to the
Haldane-phase region in the phase diagram shown in
Fig.~\ref{fig:phasediagram}.  The estimations of the limiting
(\hbox{$L\!\to\!\infty$}) values of these energy gaps have been performed again
in the same way as that in the case of the point
$(\Delta_{\rm l}\!=\!0.5,J_{{\rm l},b}\!=\!-0.5)$
(see the solid lines in Fig.~\ref{fig:gapHAL}).  Then, we have obtained the
result that the former gap vanishes [$\Delta E^{\rm O}(1)$ is even negative to
be \hbox{$\sim\!-9\!\times\!10^{-5}$}].
{
For the latter gap, we have obtained the extrapolated value
\hbox{$\Delta E^{\rm O}(2)\!\sim\!0.002$}, which is quite small but seems
to be finite.  These results indicate that the system exhibits a small but
nonzero Haldane gap.}
[Note that in the large ($L \to \infty$) system under the OBC, the
lowest-energy states in the subspace of $M=0$ and $1$ are degenerate, and
$\Delta E^{\rm O}(2)$ gives the excitation gap above the degenerate ground
states.]  The local magnetizations $m_{\ell}^{\rm O}(j;L,M)$ for
\hbox{$L\!=\!96$} and \hbox{$M\!=\!1$} at the point
$({\Delta_{\rm l}\!=\!0.25},{J_{{\rm l},b}\!=\!1.0})$ are plotted as functions
of $j$ in Fig.~\ref{fig:mHAL}.  This figure clearly demonstrates that the
edge-localized spins appear, which is one of the most distinguishing features
of the Haldane state in the $S\!=\!1$ chain.
All of these results are consistent with the conclusion of the ED analysis that
the system at the parameter point
$({\Delta_{\rm l}\!=\!0.25},{J_{{\rm l},b}\!=\!1.0})$ is in the Haldane state
with a small excitation gap.  On the other hand, our DMRG results for the
correlation functions in the system with $L = 96$, which are not shown here,
are unfortunately not able to determine their asymptotic behaviors due to the
strong open-boundary effect coming from the large correlation length.

\begin{figure}[t]
   \begin{center}
       \scalebox{0.31}{\includegraphics{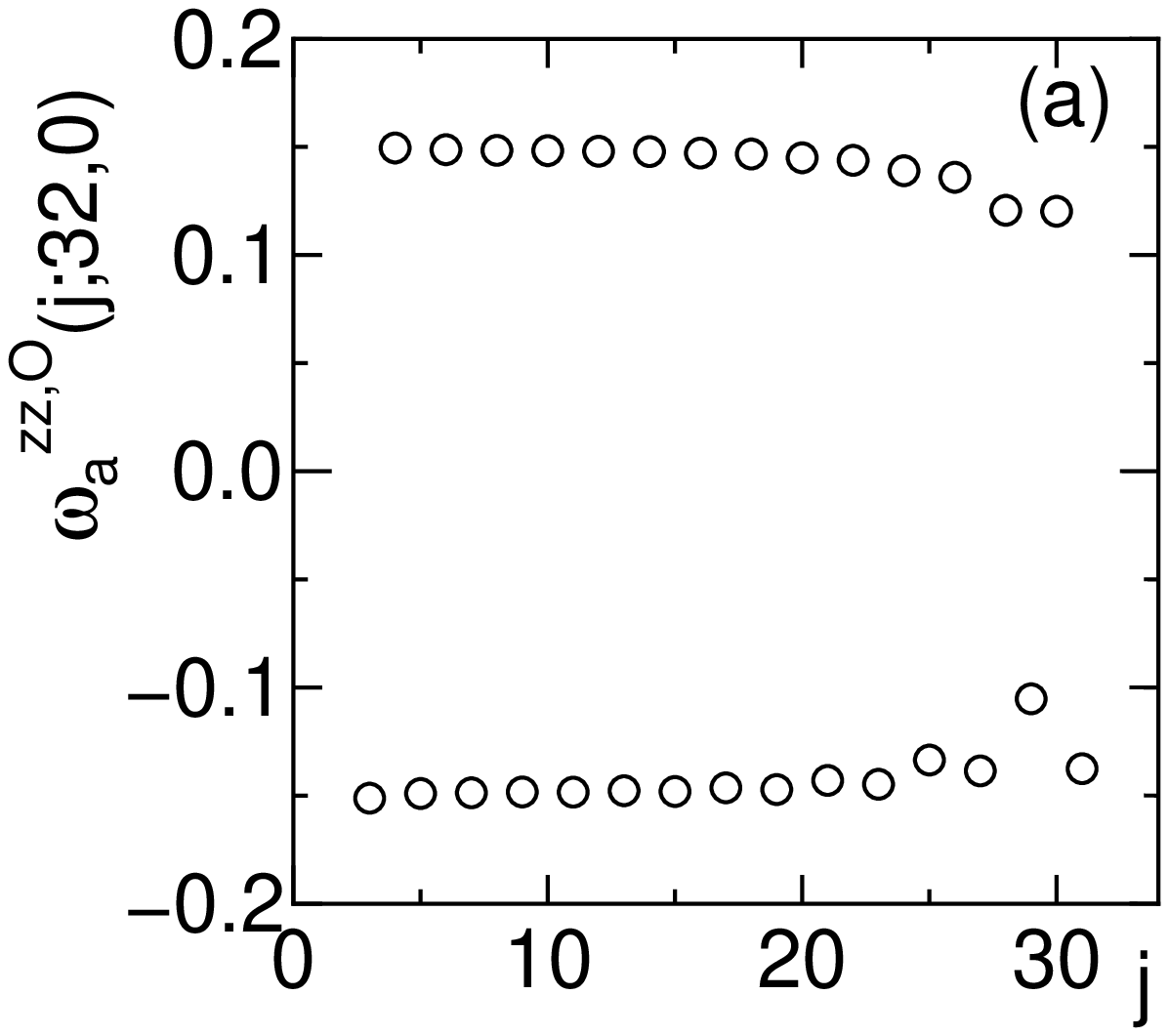}}~~
       \scalebox{0.31}{\includegraphics{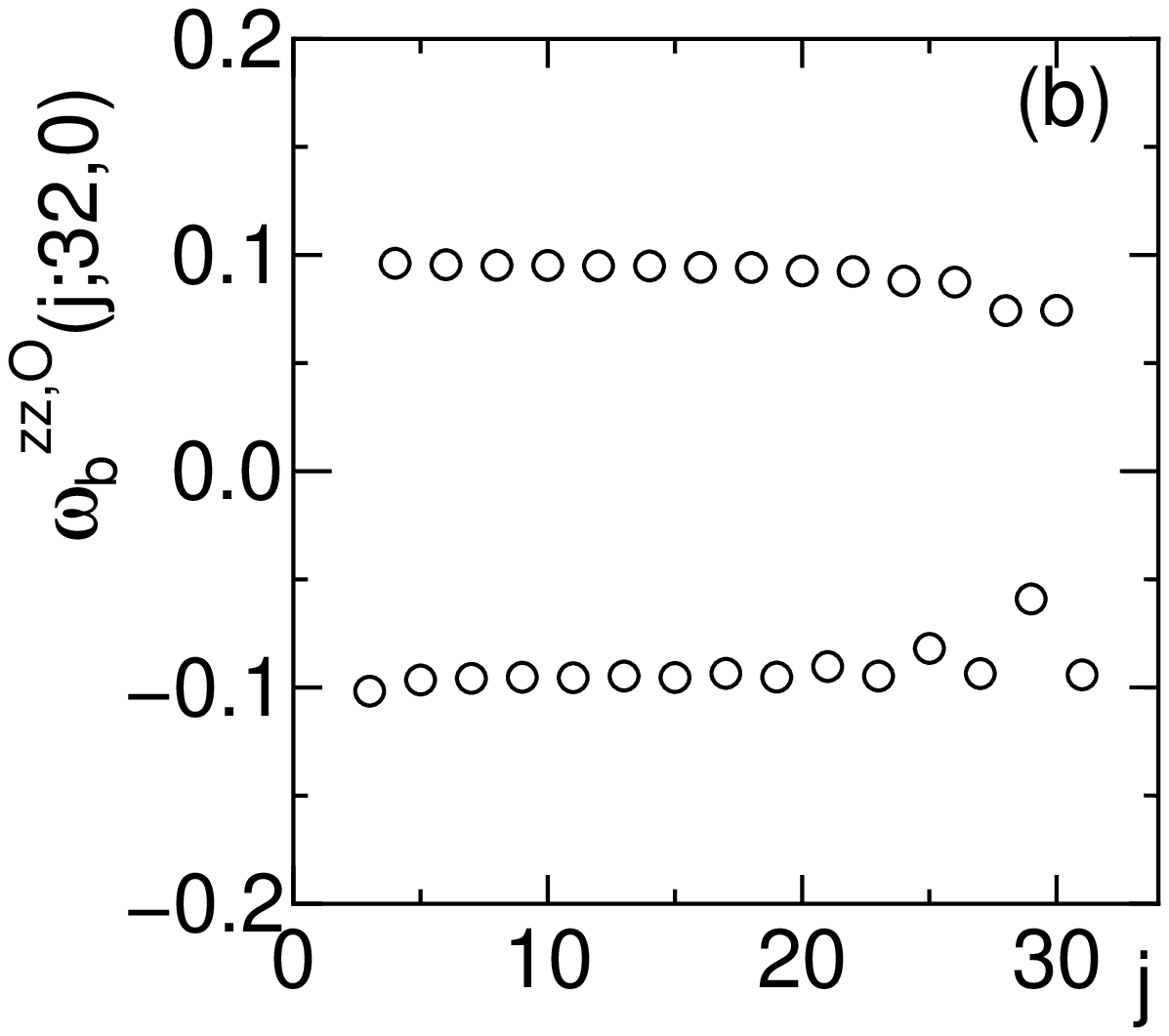}}
   \end{center}
   \caption{Dependences of (a) $\omega_{a}^{zz,{\rm O}}(j;32,0)$ and
   (b) $\omega_{b}^{zz,{\rm O}}(j;32,0)$ at the point
   $(\Delta_{\rm l}\!=\!-0.5,J_{{\rm l},b}\!=\!-1.0)$ (in the N{\'e}el phase)
   in the phase diagram shown in Fig.~\ref{fig:phasediagram}.
   }
   \label{fig:corfnNEEL1}
\end{figure}

Finally, we depict in Fig.~\ref{fig:corfnNEEL1} the $j$-dependences of the
correlation functions $\omega_{a}^{zz,{\rm O}}(j;L,M)$ and
$\omega_{b}^{zz,{\rm O}}(j;L,M)$ for \hbox{$L\!=\!32$}\cite{small-system-Neel}
and \hbox{$M\!=\!0$} at
the point $(\Delta_{\rm l}\!=\!-0.5,J_{{\rm l},b}\!=\!-1.0)$, which belongs to
the one of the N{\'e}el-phase regions in the phase diagram shown in
Fig.~\ref{fig:phasediagram}.  From this figure, we see that the strong N{\'e}el
long-range order (LRO) certainly exists, supporting our conclusion obtained by
the ED analysis that the system at this parameter point indeed belongs to the
N{\'e}el phase.  It is noted that we have also
obtained a similar result at the point
$(\Delta_{\rm l}\!=\!0.75,J_{{\rm l},b}\!=\!1.0)$ which belongs to another
N{\'e}el-phase region.

\section{Concluding Remarks}

We have numerically determined the ground-state phase diagram on the
$\Delta_{\rm l}$ versus $J_{{\rm l},b}$ plane of the \hbox{$S\!=\!1/2$} two-leg
ladder with different leg interactions, which is governed by the
Hamiltonian ${\cal H}$ of Eq.~(\ref{eq:hamiltonian}), confining ourselves to
the case where \hbox{$J_{{\rm l},a}\!=\!0.2$}, \hbox{$J_{{\rm r}}\!=\!-1.0$},
and \hbox{$\Gamma_{\rm r}\!=\!0.5$}.  The resultant phase diagram, which is
presented in Fig.~\ref{fig:phasediagram}, consists of the
ferromagnetic, Haldane, N{\'e}el, nematic TLL, partial ferrimagnetic, and
$XY1$ phases.  In addition, the phase diagram on the $\Gamma_{\rm r}$
versus $J_{{\rm l},b}$  plane in the case where \hbox{$J_{{\rm l},a}\!=\!0.2$},
\hbox{$J_{{\rm r}}\!=\!-1.0$}, and \hbox{$\Delta_{\rm l}\!=\!1.0$}, in which
the nematic TLL phase also appears, has been obtained (see
Fig.~\ref{fig:phasediagram-a} later).  By estimating the phase boundary lines,
we have successfully applied the LS~\cite{okamoto-nomura,nomura-okamoto-1,
nomura-okamoto-2, LS-NK} or PRG~\cite{PRGmethod} analysis of the numerical
data obtained by the ED method to the transition between
two of the Haldane, N{\'e}el, nematic TLL, and $XY1$ phases, while we have
compared the energies $E_0^{\rm P}(L,M)$'s for the transition associated with
the ferromagnetic and partial ferrimagnetic phases.  In our opinion, one of
the significant results of the present work is the fact that the nematic TLL
phase appears in the strong-rung unfrustrated region as well as in the
strong-rung frustrated region.

We have carried out first-order perturbational calculations from the
strong-rung
coupling limit, in which the present system~(${\cal H}$) is mapped onto the
anisotropic \hbox{$S\!=\!1$} chain system ($\cHeff$), in order to outline the
essential features of the phase diagram.  Furthermore, performing the
DMRG~\cite{dmrg-white-1,dmrg-white-2} calculations, we have clarified the
following.  In the Haldane phase, there is a finite excitation gap (the
Haldane gap) above the ground state, and the edge-localized spins appear.
In the N{\'e}el phase, the magnetic LRO of the N{\'e}el type is found.  In
the nematic TLL phase, the two-magnon bound state characterizes this phase, and
the nematic four-spin correlation function with a quasi-LRO is more dominant
than the longitudinal two-spin correlation function with a quasi-LRO
and the transverse two-spin correlation function with a short-range order.
In the $XY1$ phase, all of the three correlation functions exhibit a
quasi-LRO, and among them the transverse two-spin correlation function is most
dominant.

The recent result of the bosonization analysis for the present system,
attempted by one of the authors (S.~C.~F.) suggests that the nematic TLL
phase also appears in the weak-rung frustrated region.
This prediction is highly plausible for the following reason.  Let us consider
a two-magnon excited state in the system of almost independent chains with
$(J_{{\rm l},a},\Delta_{\rm l})$ and $(J_{{\rm l},b},\Delta_{\rm l})$ coupled
via the weak rung interaction $(J_{\rm r}, \Gamma_{\rm r})$.  We assume that
\hbox{$J_{{\rm l},a}\!>\!0.0$}, \hbox{$J_{{\rm l},b}\!<\!0.0$}, 
\hbox{$J_{\rm r}\!<\!0.0$}, \hbox{$|\Delta_{\rm l}|\!<\!1.0$}, and
\hbox{$0.0\!\le\!\Gamma_{\rm r}\!<\!1.0$}.  The ferromagnetic Ising-like rung
interaction
favors the formation of the two-magnon bound state in a rung, as mentioned in
Sect.~1.  On the other hand, if the one-magnon excitation energies in each
chain are different, the leg interactions result in two magnons sitting in the
chain with the smaller excitation energy, resulting in a scattered state of two
magnons.  Therefore, at the level of a crude approximation, one can expect that
the two magnons form a bound state when $|J_r|$ is sufficiently larger than the
difference between the one-magnon excitation energies in each chain,
\begin{eqnarray}
   \!\!\!\!\!\!\!\!\!\!\!\!\!\!\!\!\!\!\!\!&&\!\!\!\!\!
          |\Delta(\Delta E)_{\rm om}|                          \nonumber \\
   \!\!\!\!\!\!\!\!\!\!\!\!\!\!\!\!\!\!\!\!&&
           =|(\Delta E)_{\rm om}(L;J_{{\rm l},a},\Delta_{\rm l})
                   - (\Delta E)_{\rm om}(L;J_{{\rm l},b},\Delta_{\rm l})|\,,
\label{eq:dif_exc_one_magnon}
\end{eqnarray}
where $(\Delta E)_{\rm om}(L;J_\ell,\Delta_{\rm l})$ is the one-magnon
excitation energy in an independent chain with
$J_{{\rm l},\ell}$ and $\Delta_{\rm l}$.  The \hbox{$L\to\infty$} asymptotic
form of the one-magnon excitation energy is obtained from the conformal field
theory~\cite{cf-theory} and the Bethe ansatz~\cite{B-ansatz-1,B-ansatz-2} as
\begin{eqnarray}
   \!\!\!\!\!\!\!\!\!\!\!\!\!\!\!\!\!\!\!\!&&
       (\Delta E)_{\rm om}(L;J_{{\rm l},a},\Delta_{\rm l})
           \sim \frac{\pi}{L}\; v(J_{{\rm l},a},\Delta_{\rm l}) \;
           \eta (\Delta_{\rm l})\,,
\label{eq:one-magnon_exc_lega} \\
   \!\!\!\!\!\!\!\!\!\!\!\!\!\!\!\!\!\!\!\!&&
       (\Delta E)_{\rm om}(L;J_{{\rm l},b},\Delta_{\rm l})
           \sim \frac{\pi}{L}\; v(|J_{{\rm l},b}|,-\Delta_{\rm l}) \;
           \eta (-\Delta_{\rm l})
\label{eq:one-magnon_exc_legb}
\end{eqnarray}
with
\begin{eqnarray}
    v(J,\Delta) &=& \frac{\pi \sqrt{1-\Delta^2}}{2 \cos^{-1}\Delta}\; J  \,,
\label{eq:velocity_Bethe} \\
    \eta(\Delta) &=& 1- \frac{1}{\pi} \cos^{-1}\Delta\,.
\label{eq:eta_Bethe}
\end{eqnarray}
Note that the chain with $(J_{{\rm l},b},\Delta_{\rm l})$ is transformed into
the chain with $(|J_{{\rm l},b}|,-\Delta_{\rm l})$ by the $\pi$-rotation around
the $z$ axis for every other spin.  Thus, solving the equation
\hbox{$|\Delta(\Delta E)_{\rm om}|\!=\!0$} with
Eqs.~(\ref{eq:one-magnon_exc_lega})-(\ref{eq:eta_Bethe}), we can determine the
value $\tilde\Delta_{\rm l}$ of $\Delta_{\rm l}$ at which the difference in the
one-magnon excitation energies vanishes.  The result is
\begin{equation}
   \tilde\Delta_{\rm l}
      = \cos \Bigg(\frac {\pi \sqrt{J_{{\rm l},a}/|J_{{\rm l},b}|}}
                          {1+\sqrt{J_{{\rm l},a}/|J_{{\rm l},b}|}}\Bigg)\,,
\end{equation}
from which, when \hbox{$J_{{\rm l},a}\!=\!0.2$}, for example,
\hbox{$\tilde\Delta_{\rm l}\!=\!0.0$} for \hbox{$J_{{\rm l},b}\!=\!-0.2$} and
\hbox{$\tilde\Delta_{\rm l}\!=\!0.158$} for \hbox{$J_{{\rm l},b}\!=\!-0.3$}.
We thereby expect that, at least for a parameter region around the line
$\Delta_{\rm l} = \tilde\Delta_{\rm l}$, the inclusion of weak ferromagnetic,
Ising-like rung interactions of $J_r$ leads to the formation of a two-magnon
bound state.  Here, we should mention that the above argument is not sufficient
for one-dimensional systems with strong quantum fluctuations, and more
sophisticated approaches such as the weak-coupling renormalization group method
are required.  We are now investigating the ground-state phase diagram in this
weak-rung region, and the results will be reported elsewhere.

Finally, we hope that the present work will stimulate future experimental
studies on related subjects, which include the synthesis of a ladder
system with different leg interactions.

\acknowledgments
We acknowledge Professor Kiyohide Nomura for useful comments.
This work has been partly supported by JSPS KAKENHI Grant Numbers JP15K05198,
JP15K05882 (J-Physics), JP16J04731, JP16K05419, JP17H02931, and 
P18H04330 (J-Physics).  We also thank the Supercomputer Center, Institute for
Solid State Physics, University of Tokyo and the Computer Room, Yukawa
Institute for Theoretical Physics, Kyoto University for computational
facilities.

\section*{Appendix}

\makeatletter
\def\theequation{A$\cdot$\arabic{equation}}
\makeatother
\setcounter{equation}{0}
\makeatletter
\def\thefigure{A$\cdot$\arabic{figure}}
\makeatother
\setcounter{figure}{0}

Figure~\ref{fig:phasediagram-a} demonstrates the obtained ground-state phase
diagram on the $\Gamma_{\rm r}$ versus $J_{{\rm l},b}$ plane in the case where
\hbox{$J_{{\rm l},a}\!=\!0.2$}, \hbox{$J_{{\rm r}}\!=\!-1.0$}, and
\hbox{$\Delta_{\rm l}\!=\!1.0$}.  This consists of the N{\'e}el, Haldane,
nematic TLL, $XY$1, ferromagnetic, and partial ferrimagnetic phases, all of
which are the same as those in Fig.~\ref{fig:phasediagram}.
In this phase diagram, the phase boundary line between the N{\'e}el and $XY$1
phases is drawn by the blue line, and the colors of the other lines have the
same meanings as those in Fig.~\ref{fig:phasediagram}.
\begin{figure}[t]
   \begin{center}
       \scalebox{0.45}{\includegraphics{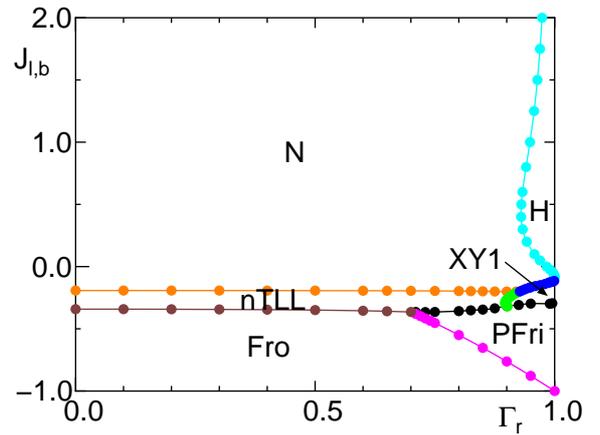}}
   \end{center}
   \caption{(Color online)
   The ground-state phase diagram on the $\Gamma_{\rm r}$ versus
   $J_{{\rm l},b}$ plane in the case where \hbox{$J_{{\rm l},a}\!=\!0.2$},
   \hbox{$J_{{\rm r}}\!=\!-1.0$}, and \hbox{$\Delta_{\rm l}\!=\!1.0$},
   obtained in the present work.  The regions designated by N, H, nTLL, XY1,
   Fro, and PFri are those of the N{\'e}el, Haldane, nematic TLL, $XY$1,
   ferromagnetic, and partial ferrimagnetic phases, respectively.
   }
   \label{fig:phasediagram-a}
\end{figure}

All of the phase boundary lines, except for that for the first-order phase
transitions between the partial ferrimagnetic phase and one of the nematic TLL
and $XY$1 phases, have been numerically determined in a way similar to those
discussed in Sect.~2; in this case, Okamoto and Nomura's LS
method,~\cite{okamoto-nomura,nomura-okamoto-1,nomura-okamoto-2} where we use
Eq.~(\ref{eq:N-XY1}), works very well for
estimating the phase boundary line between the N{\'e}el and $XY$1 phases.
Obtaining the phase boundary line for the above first-order phase transitions,
we have calculated $M_{\rm g}(L)$ by carrying out DMRG calculations for the
finite system with \hbox{$L\!=\!36$} rungs under the OBC.  We consider that the
result for the \hbox{$L\!=\!36$} system gives a fairly good approximation of
the result at the \hbox{$L\!\to\!\infty$} limit.

\begin{figure}[t]
   \begin{center}
       \scalebox{0.25}{\includegraphics{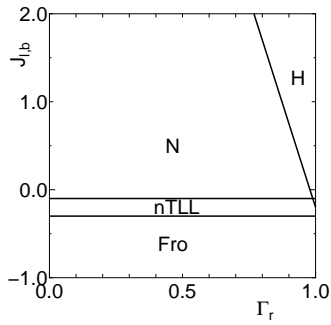}}~~
   \end{center}
   \caption{Schematic phase diagram of the present system (${\cal H}$)
    predicted by the perturbation theory based on $\cHeff$ in the same case
    (\hbox{$J_{{\rm l},a}\!=\!0.2$}, \hbox{$J_\rr\!=\!-1.0$}, and
    \hbox{$\Delta_\rl\!=\!1.0$}) as that in Fig~.\ref{fig:phasediagram-a}.
    The regions designated by N, H, nTLL, and Fro are those of the N{\'e}el,
    Haldane, nematic TLL, and ferromagnetic phases, respectively.
    }
   \label{fig:mapping2}
\end{figure}
The perturbation theory developed in Sect.~3 predicts the phase diagram shown
in Fig.~\ref{fig:mapping2} in the following way.  Here, the parameter
$\Deltaeff$ is given by \hbox{$\Deltaeff\!=\!1.0$}, since
\hbox{$\Delta_\rl\!=\!1.0$}.  Thus, the \hbox{$J_{\rl,b}\!>\!-0.2$} and
\hbox{$J_{\rl,b}\!<\!-0.2$} regions in Fig.~\ref{fig:phasediagram-a}
correspond,
respectively, to the downward- and upward-arrowed blue solid lines in
Fig.~\ref{fig:mapping}(a).  The Haldane-N{\'e}el transition point on the
\hbox{$\Deltaeff\!=\!1.0$} line in Fig.~\ref{fig:mapping}(a) can be read as
\hbox{$\Deff \simeq -0.21$}, which leads to
\begin{equation}
    J_{\rl,b} = -9.5\Gamma_\rr + 9.3\,,
\end{equation}
by using Eqs.~(20) and (22).  This line well explains the Haldane-N{\'e}el
boundary line in Fig.~\ref{fig:phasediagram-a} near the point
\hbox{$(\Gamma_\rr,J_{l,b})\!=\!(1.0,0.0)$}, around which the perturbation
theory is developed.  The crossing of the blue dotted line with the bandlike
nematic TLL region in Fig.~\ref{fig:mapping}(a) leads to the horizontal nematic
TLL band in Fig.~\ref{fig:mapping2}.  The predicted phase diagram given in
Fig.~\ref{fig:mapping2} qualitatively explains the numerical one given in
Fig.~\ref{fig:phasediagram-a}, although the $XY1$ and partial ferrimagnetic
phases do not appear in Fig.~\ref{fig:mapping2}.
{
The appearance of these phases in Fig.~\ref{fig:phasediagram-a}
is due to the frustration effect and is beyond the description of the
first-order perturbation theory.
We note that the existence of the $XY1$ phase is considered to be an
example of the inversion phenomenon concerning the interaction
anisotropy~\cite{inv-1,inv-2,inv-3,inv-4} stated in Sect.~1.}


\begin{thebibliography}{9}

\bibitem{majumdar-ghosh} C. K. Majumdar and D. K. Ghosh, J. Math. Phys.
{\bf 10}, 1399 (1969).

\bibitem{tonegawa-harada} T. Tonegawa and I. Harada, J. Phys. Soc. Jpn.
{\bf 56}, 2153 (1987).

\bibitem{okamoto-nomura} K. Okamoto and K. Nomura, Phys. Lett. A {\bf 169},
433 (1992).

\bibitem{tonegawa-harada-kaburagi} T. Tonegawa, I. Harada, and M. Kaburagi,
J. Phys. Soc. Jpn. {\bf 61}, 4665 (1992).

\bibitem{nomura-okamoto-1} K. Nomura and K. Okamoto, J. Phys. Soc. Jpn.
{\bf 62}, 1123 (1993).

\bibitem{nomura-okamoto-2} K. Nomura and K. Okamoto, J. Phys. A {\bf 27},
5773 (1994).

\bibitem{schulz} H. J. Schulz, Phys. Rev. B {\bf 34}, 6372 (1986)

\bibitem{nijs-rommelse} M. den Nijs and K. Rommelse,
Phys. Rev. B {\bf 40}, 4709 (1989).

\bibitem{chen-etal} W. Chen, K. Hida, and B. C. Sanctuary, Phys. Rev. B
{\bf 67}, 104401 (2003).

\bibitem{lauchli-etal} A. L{\"a}uchli, G. Schmid, and S. Trebst, Phys. Rev.
B {\bf 74}, 144426 (2006).

\bibitem{vekua-etal} T. Vekua, A. Honecker, H.-J. Mikeska, and
F. Heidrich-Meisner, Phys. Rev. B {\bf 76}, 174420 (2007).

\bibitem{hikihara-etal-1} T. Hikihara, L. Kecke, T. Momoi, and A. Furusaki,
Phys. Rev. B {\bf 78}, 144404 (2008).

\bibitem{sudan-etal} J. Sudan, A. L{\"u}scher, and A. M. L{\"a}uchli, Phys.
Rev. B {\bf 80}, 140402(R) (2009).

\bibitem{frustrated-ladder-1} A. Lavar{\'e}lo, G. Guillaume, and
N. Laflorencie, Phys. Rev. B {\bf 84}, 144407 (2011), and references therein.

\bibitem{frustrated-ladder-2} T. Vekua and A. Honecker, Phys. Rev. B {\bf 73},
214427 (2006), and references therein.

\bibitem{frustrated-ladder-3} F. Michaud, T. Coletta, S. R. Manmana,
J.-D. Picon, and F. Mila, Phys. Rev. B {\bf 81}, 014407 (2010), and references
therein.

\bibitem{ladder-altrung-1} G. I. Japaridze and E. Pogosyan, J. Phys.: Condens.
Matter {\bf 18}, 9297 (2006).

\bibitem{ladder-altrung-2} F. Amiri, G. Sun, H.-J. Mikeska, and T. Vekua, Phys.
Rev. B {\bf 92}, 184421 (2015).

\bibitem{ladder-altrung-tone} T. Tonegawa, K. Okamoto, T. Hikihara, and
T. Sakai, J. Phys.: Conf. Ser. {\bf 683}, 012039 (2016).

\bibitem{inv-1} K. Okamoto and Y. Ichikawa, J. Phys. Chem. Solids {\bf 63},
1575 (2002).

\bibitem{inv-2} K. Okamoto, Prog. Theor. Phys. Suppl. {\bf 145}, 208 (2002).

\bibitem{inv-3} A. Tokuno and K. Okamoto J. Phys. Soc. Jpn. Suppl. {\bf 74},
157 (2005).

\bibitem{inv-4} K. Okamoto, JPS Conf. Proc. {\bf 1}, 012031 (2014).

\bibitem{yamaguchi-etal-1} H. Yamaguchi, K. Iwase, T. Ono, T. Shimokawa,
H. Nakano, Y. Shimura, N. Kase, S. Kittaka, T. Sakakibara, T. Kawakami,
and Y. Hosokoshi, Phys. Rev. Lett. {\bf 110}, 157205 (2013).

\bibitem{yamaguchi-etal-2} H. Yamaguchi, H. Miyagai, T. Shimokawa, K. Iwase,
T. Ono, Y. Kono, N. Kase, K. Araki, S. Kittaka, T. Sakakibara, T. Kawakami,
K. Okunishi, and Y. Hosokoshi, J. Phys. Soc. Jpn. {\bf 83}, 033707 (2014).

\bibitem{OkunishiT2003}
K.\ Okunishi and T.\ Tonegawa, J. Phys. Soc. Jpn. {\bf 72}, 479 (2003).

\bibitem{HikiharaMFK2010}
T.\ Hikihara, T.\ Momoi, A.\ Furusaki, and H.\ Kawamura, Phys. Rev. B {\bf 81}, 224433 (2010).

\bibitem{tsukano-takahashi} M. Tsukano and M. Takahashi, J. Phys. Soc. Jpn.
{\bf 66}, 1153 (1997).

\bibitem{ladder-diffleg-tone} T. Tonegawa, K. Okamoto, T. Hikihara, and
T. Sakai, J. Phys.: Conf. Ser. {\bf 828}, 012003 (2017).

\bibitem{sekiguchi-hida} K. Sekiguchi and K. Hida, J. Phys. Soc. Jpn.
{\bf 86}, 084706 (2017).

\bibitem{hikihara-etal-2} T. Hikihara, T. Tonegawa, K. Okamoto, and T. Sakai,
J. Phys. Soc. Jpn. {\bf 86}, 054709 (2017).

\bibitem{dmrg-white-1}
S. R. White, Phys. Rev. Lett. {\bf 69}, 2863 (1992).

\bibitem{dmrg-white-2}
S. R. White, Phys. Rev. B {\bf 48}, 10345 (1993).

\bibitem{BKT-1} Z. L. Berezinskii, Sov.\ Phys. JETP {\bf 34}, 610 (1971).
                            
\bibitem{BKT-2} J. M. Kosterlitz and D. J. Thouless, J. Phys. C {\bf 6},
1181 (1973).

\bibitem{LS-NK} K. Nomura and A. Kitazawa, J. Phys. A {\bf 31}, 7341 (1998).

\bibitem{PRGmethod}
M. P. Nightingale, Physica A {\bf 83}, 561 (1976).

\bibitem{four-rung-period}
The four-rung-period oscillation in $\omega_{\ell}^{zz,{\rm O}}(j;96,0)$ found
in Fig.~\ref{fig:corfnNTLL} can be understood as follows.  Let $d$ ($d'$) be
the distance between the $j_0$th [($j_0+j$)th] rung and the left (right) edge
of the open ladder. Then, as $j$ increases, the parity of $(d, d')$ changes
as $\cdots$, (even, even), (even, odd), (odd, odd), (odd, even), $\cdots$. 

\bibitem{bosonization-text}
T.\ Giamarchi, {\it Quantum Physics in One Dimension} (Oxford University Press,
New York, 2004).

\bibitem{small-system-Neel}
The reason why we show the results for a rather small system with
\hbox{$L\!=\!32$} here is as follows.  In
the N{\'e}el phase, the finite system has a unique ground state,
which is represented by a bonding or antibonding state of the two lowest-energy
states in the subspace of $M=0$, and the local magnetization is zero, namely,
\hbox{$m_{\ell}^{\rm O}(j;L,M)\!=\!0$}.  However, in the DMRG calculation, one
of the two lowest-energy states is selected (depending on a randomly prepared
initial state) for a finite but large system, resulting in the nonzero local
magnetization \hbox{$m_{\ell}^{\rm O}(j;L,M)\!\ne\!0$} even for
\hbox{$M\!=\!0$}.  Indeed, we have obtained such a ground state with nonzero
$m_{\ell}^{\rm O}(j;L,M)$ for \hbox{$L\!\ge\!48$}.  Although this observation
of \hbox{$m_{\ell}^{\rm O}(j;L,M)\!\ne\!0$} for large systems can also be
regarded as an indication of the N{\'e}el phase, we have restricted ourselves
to treating the rather small systems with \hbox{$L\!\le\!32$}, which preserve
\hbox{$m_{\ell}^{\rm O}(j;L,M)\!=\!0$}.

\bibitem{cf-theory} J. H. Cardy, J. Phys. A {\bf 17}, L385 (1984).

\bibitem{B-ansatz-1} J. Des Cloizeaux and M. Gaudin, J. Math. Phys. {\bf 7},
1384 (1966).

\bibitem{B-ansatz-2} J. D. Johnson, S. Krinsky, and B. McCoy, Phys. Rev. A
{\bf 8}, 2526 (1973).

\end{thebibliography}
\end{document}